\documentclass[twocolumn,preprintnumbers]{aastex6} 
\usepackage{afterpage}
\usepackage{capt-of}
\usepackage{diagbox}

\usepackage[figuresright]{rotating}
\usepackage{amsmath}
\usepackage{amssymb,amsmath,latexsym,graphics, graphicx,epsfig,multirow,comment,hyperref,appendix,feyn,slashed,xcolor,afterpage, makecell} 
\usepackage{booktabs}
\usepackage{tabularx}

\newcommand{\FeH}{\text{[Fe/H]} }

\newcommand {\be} {\begin {equation}}
\newcommand {\ee} {\end {equation}} 

\newcommand {\bes} {\begin {equation*}}
\newcommand {\ees} {\end {equation*}}

\newcolumntype{L}[1]{>{\raggedright\let\newline\\\arraybackslash\hspace{0pt}}m{#1}}
\newcolumntype{C}[1]{>{\centering\let\newline\\\arraybackslash\hspace{0pt}}m{#1}}
\newcolumntype{R}[1]{>{\raggedleft\let\newline\\\arraybackslash\hspace{0pt}}m{#1}}

\usepackage{csvsimple}

\makeatletter
\newcommand\footnoteref[1]{\protected@xdef\@thefnmark{\ref{#1}}\@footnotemark}
\makeatother

\DeclareRobustCommand{\Fig}[1]{Fig.~\ref{#1}}

\DeclareRobustCommand{\Eq}[1]{Eq.~(\ref{#1})}

\newcommand{\beq}{\begin{equation}}
\newcommand{\eeq}{\end{equation}}

\begin{document}

\title{Inferred Evidence For Dark Matter Kinematic Substructure with SDSS-\emph{Gaia}}
\preprint{}

\author{Lina Necib}
\affil{Walter Burke Institute for Theoretical Physics,
California Institute of Technology, Pasadena, CA 91125, USA}

\author{Mariangela Lisanti}
\affil{Department of Physics, Princeton University, Princeton, NJ 08544, USA}

\author{Vasily Belokurov}
\affil{Institute of Astronomy, University of Cambridge, Madingley Road, Cambridge CB3 0HA, UK}
\affil{Center for Computational Astrophysics, Flatiron Institute, 162 5th Avenue, New York, NY 10010, USA}

\begin{abstract}

We use the distribution of accreted stars in SDSS-\emph{Gaia} DR2 to demonstrate that a non-trivial fraction of the dark matter halo within Galactocentric radii of 7.5--10~kpc and $|z| > 2.5$~kpc is in substructure, and thus may not be in equilibrium.
Using a mixture likelihood analysis, we separate the contributions of an old, isotropic stellar halo and a younger anisotropic population.  The latter dominates and is uniform within the region studied. It can be explained as the tidal debris of a disrupted massive satellite on a highly radial orbit, and is consistent with mounting evidence from recent studies.  
Simulations that track the tidal debris from such mergers find that the dark matter traces the kinematics of its stellar counterpart.  If so, our results indicate that a component of the nearby dark matter halo that is sourced by luminous satellites is in kinematic substructure referred to as debris flow.  These results challenge the Standard Halo Model, which is discrepant with the distribution recovered from the stellar data, and have important ramifications for the interpretation of direct detection experiments.

\end{abstract}
\maketitle
\pagebreak

\section{Introduction} 
\label{sec:intro}

The motions of nearby stars have proven invaluable in the study of the local dark matter (DM) distribution, starting from the work of Kapteyn~\citep{1922ApJ....55..302K}, Jeans~\citep{1922MNRAS..82..122J},  and Oort~\citep{1932BAN.....6..249O}.  As the mapping of stellar velocities has dramatically improved with surveys such as Hipparcos~\citep{1997ESASP1200.....E}, the Sloan Digital Sky Survey (SDSS)~\citep{2012ApJS..203...21A}, and the RAdial Velocity Experiment~\citep{2017AJ....153...75K},  the measurement
of the local DM density has greatly improved~\citep{Read:2014qva}.  In contrast, the DM velocity distribution has remained relatively unexplored.  In this work, we use data from the second \emph{Gaia} data release~\citep{2016A&A...595A...1G, 2018arXiv180409365G}, cross-matched with SDSS, to characterize the local population of accreted stars and infer properties of the DM velocity distribution.  

In the $\Lambda$CDM paradigm, the Milky Way halo is built from the mergers of smaller satellite galaxies~\citep{White:1977jf}.  As these galaxies fall into the Milky Way, they experience strong tidal forces that gradually tear them apart.  In the initial stages of disruption, the DM in the outskirts of the satellite is stripped away, but as the galaxy is slowly eaten down, its more tightly bound stars and DM are also removed.  This tidal debris litters the Milky Way, a fossil remnant of the Galaxy's accretion history.  

DM that accreted early has time to virialize with the host galaxy.  We can thus imagine that it forms an isotropic isothermal halo~\citep{1974ApJ...193L...1O, 1980ApJS...44...73B, Caldwell:1981rj} with a mass distribution that is consistent with a flat rotation curve~\citep{1979ApJ...231L.115B,1978A&A....63....7B,1985ApJ...295..422C,1985AJ.....90..254K,1989ApJ...342..272F,1994A&A...285..415P}.  Self-consistently, the local DM velocities follow a Maxwell-Boltzmann distribution~\citep{Drukier:1986tm, Freese:1987wu}.  This scenario is often referred to as the Standard Halo Model.  
 
More recent mergers, however, can leave residual structures in the DM phase-space distribution.  For example, the debris from the most recent mergers may be in cold phase-space structures called tidal streams, which are coherent in position and velocity space~\citep{Zemp:2008gw, Vogelsberger:2008qb, 2008Natur.454..735D,Kuhlen:2009vh,Maciejewski:2010gz,2011MNRAS.413.1419V, Elahi:2011dy}.  With time, the tidal debris becomes increasingly more mixed.  In this process, any velocity features  typically persist longer than spatial ones~\citep{Helmi:1999ks}.  The resulting kinematic substructure, referred to as debris flow, may consist of many overlapping streams, shells or plumes from the debris of one or more satellites that have made several orbits before dissolving~\citep{Lisanti:2011as, Kuhlen:2012fz}.   

Numerical simulations are an invaluable tool in exploring the range of DM distributions possible in Milky Way--like galaxies.  They have demonstrated, for example, that the Solar neighborhood could have been built from a wide variety of accretion histories~\citep{Diemand:2007qr,2011MNRAS.413.1373W}.  DM-only simulations find fairly marked deviations from the Maxwell-Boltzmann assumption, especially on the high-end tail of the velocity distribution~\citep{Vogelsberger:2008qb,MarchRussell:2008dy,Fairbairn:2008gz,Kuhlen:2009vh,Mao:2012hf}.  Simulations that also model the baryonic physics typically exhibit closer alignment with the Maxwell Boltzmann, although significant scatter is observed between them~\citep{Ling:2009eh,Pillepich:2014784,Bozorgnia:2016ogo, Kelso:2016qqj, Sloane:2016kyi, Bozorgnia:2017brl}.

Ultimately, we would like to determine the local DM distribution from observations.  One possibility is to use the motion of stars to constrain the local gravitational potential (or density) and to subsequently infer the velocity distribution using Jeans Theorem.  A variety of proposals of this nature have been made~\citep{Hansen:2004qs,Chaudhury:2010hj, Lisanti:2010qx, Bhattacharjee:2012xm,  2012JCAP...05..005C, 2013JCAP...12..050B,Fornasa:2013iaa, Mandal:2018efq}, but they typically rely on the assumption that the DM is isotropic and/or in equilibrium, either of which may be violated depending on the Milky Way's accretion history.

An alternative proposal is to identify populations of stars that share the same kinematics as the DM.  These are stars that were also accreted onto the Milky Way from merging satellites.  As such, they typically have distinctive kinematic and chemical properties compared to the population that is born in the Galaxy.  
Using accreted stars as direct kinematic tracers for the DM is beneficial because it makes no assumption about steady state.  The potential downside to this approach is that it does not account for DM that originates from non-luminous satellites or that was accreted diffusely.

Numerical simulations have demonstrated excellent correspondence between the DM and accreted stars. \cite{eris_paper} recently showed that the oldest and most metal-poor stars in the halo trace the velocities of the virialized DM using the \texttt{Eris} hydrodynamic simulation. Stellar substructure in the form of debris flow  also traces similar kinematic features in the DM distribution, as was demonstrated using the~\texttt{Via Lactea} simulation~\citep{Lisanti:2014dva}.  As this substructure arises from more recent accretion events than the virialized component, it may be associated with more metal-rich stars.  More recently, the correspondence between the stellar and DM velocities has also been explicitly demonstrated (for the virialized component and debris flow at redshift $z=0$) in two additional simulated halos in the \textsc{Fire} suite~\citep{Necib:2018}.

These arguments motivate a close study of the local accreted stellar population as a means of characterizing the DM distribution.  Recent observations have begun to change our understanding of the stellar halo, disfavoring the viewpoint that a large fraction was born \emph{in-situ} from stars that were kicked up from the Galactic disk~\citep{2008A&ARv..15..145H}.  In contrast, evidence has been building for a two-component model that consists of an isotropic population from old accretion events, and an anisotropic population from a more recent---and quite significant---merger~\citep{2018MNRAS.477.1472B, 2018arXiv180510288D, Myeong:2018kfh, 2018arXiv180606038H, 2018ApJ...856L..26M, LancasterBHBs}.  In this picture, the majority of the local stellar halo originates from this one merger, which also deposits DM.  This work presents the first modeling of the velocity and metallicities of the stars accreted in this event, providing clues about the corresponding DM debris as well.

The second \emph{Gaia} data release (DR2) presents a unique opportunity to study the kinematics of this accreted population.  Cross-matching \emph{Gaia} DR2 with SDSS yields metallicities for 193,162 of its stars.  This dataset allows us to characterize the velocity distribution of 21,443 stars with metallicities down to $\FeH \sim -3$ between Galactocentric radii of 7.5--10~kpc and $2.5 ~\rm{kpc} < |z| <10$ kpc. We build a model that successfully describes the observations over the full range of metallicities and velocities.  As we will argue, the properties of the local stellar distribution suggest that a fraction of the local DM is in substructure called debris flow, challenging the Standard Halo Model. 

Characterizing the DM velocity distribution is critical for interpreting results of direct detection experiments, which search for the recoils of nuclei from a collision with a DM particle---see~\cite{1996PhR...267..195J, Freese:2012xd} for reviews.  The rate of such scattering interactions depends on the DM speed upon collision.  Indeed, for certain DM speeds and/or models the velocity distribution can make the difference between observing a signal or seeing nothing at all.  The characterization of the DM velocity distribution function is one of the largest sources of uncertainty in the interpretation of such experimental results~\citep{Cremonesi:2013bma,Green:2017odb}. If a subset of the local DM is indeed in substructure, it could potentially change the landscape of exclusion limits on DM masses and scattering cross sections. We however caution that the analysis performed in this paper only holds for $|z|>2.5$ kpc. While we extrapolate the results to the Sun's position for illustration, a dedicated study is needed to confirm the behavior closer to the mid-plane.

Our paper is organized as follows. Sec.~\ref{sec:methodology} describes the data preparation and likelihood procedure used in the study.  Sec.~\ref{sec:stellarhalo} presents the results of the SDSS-\emph{Gaia}~DR2 analysis, and describes the characteristics of the disk, halo, and substructure stars that we identify.  Sec.~\ref{sec:darkmatter} discusses the implications of this stellar substructure for DM detection.  Figs.~\ref{fig:heliocentric} and~\ref{fig:gvmin} summarize the impact on experimental limits for  spin-independent interactions.  We conclude in Sec.~\ref{sec:conclusions}.  The Appendix includes supplementary material that further substantiates the results presented in the main text.  Interpolations of the heliocentric velocity distribution, which can be used to calculate DM scattering rates, are provided at the following \texttt{github} repository \url{https://linoush.github.io/DM_Velocity_Distribution/}.

\section{Analysis Methodology}
\label{sec:methodology}

\subsection{Data Selection}

In this analysis, we use the SDSS DR9~\citep{2012ApJS..203...21A}
dataset, cross-matched to \emph{Gaia} DR2~\citep{2018arXiv180409365G}.
The cross-match was performed inside the Whole Sky Data Base (WSDB),
an archive providing SQL access to catalogs from all major wide-area
surveys.  In particular, we utilized \texttt{Q3C}, spatial indexing
and cross-matching plug-in \citep[][]{q3c}  to select the nearest SDSS
neighbor for each \emph{Gaia} source within 1$''$ aperture, while
taking into account the proper motion of all objects as well as the
time difference between the observations. The combination of the two
datasets allows us to take advantage of the large number of halo stars
in the SDSS spectroscopic dataset, while simultaneously using the
unprecedented accuracies of the proper motions provided by the
\emph{Gaia} survey.  From the SDSS catalog, we select Main Sequence (MS) stars, which are considered standard candles in this context,
 that satisfy: $|b| > 10^\circ$, $A_g < 0.5$~mag,
$\sigma_\mathrm{RV} < 50$~km/s, $\mathrm{S/N} > 10$, $3.5 < \log(g) <
5$, $0.2 < g-r < 0.8$, $0.2 < g- i < 4$, $4500 < T_\mathrm{eff} <
8000$~K, and $15 < r < 19.5$.  All stellar magnitudes are dereddened
using the maps of \cite{Schlegel:1997yv}.
Admittedly, the MS stars are not true standard candles but accurate distances can be estimated using the so-called photometric parallax method, which was recently tested against \emph{Gaia} parallaxes (see~\citet{2018arXiv180510288D}). While other possible choices exist, \emph{e.g.} (i) a mix of all stellar populations with \emph{Gaia} parallaxes or  (ii) red giants with photometric parallaxes and (iii) blue horizontal branch stars, the MS sample has a combination of advantageous properties making it a better choice for the analysis presented here. More precisely, the MS sample is not limited by \emph{Gaia}'s parallax error to a relatively short distance range and is not dominated by the disk stars. The photometric parallax uncertainties for the MS stars do not vary as strongly with metallicity as \emph{e.g.} those for red giants (whose distance error can easily reach 100\% for the metal poor giants compared to the typical 10-20\% for the MS stars). Finally, MS starts are much more numerous compared to the blue horizontal branch or red giant stars.

\begin{figure}[t] 
\centering
\includegraphics[width=0.5\textwidth]{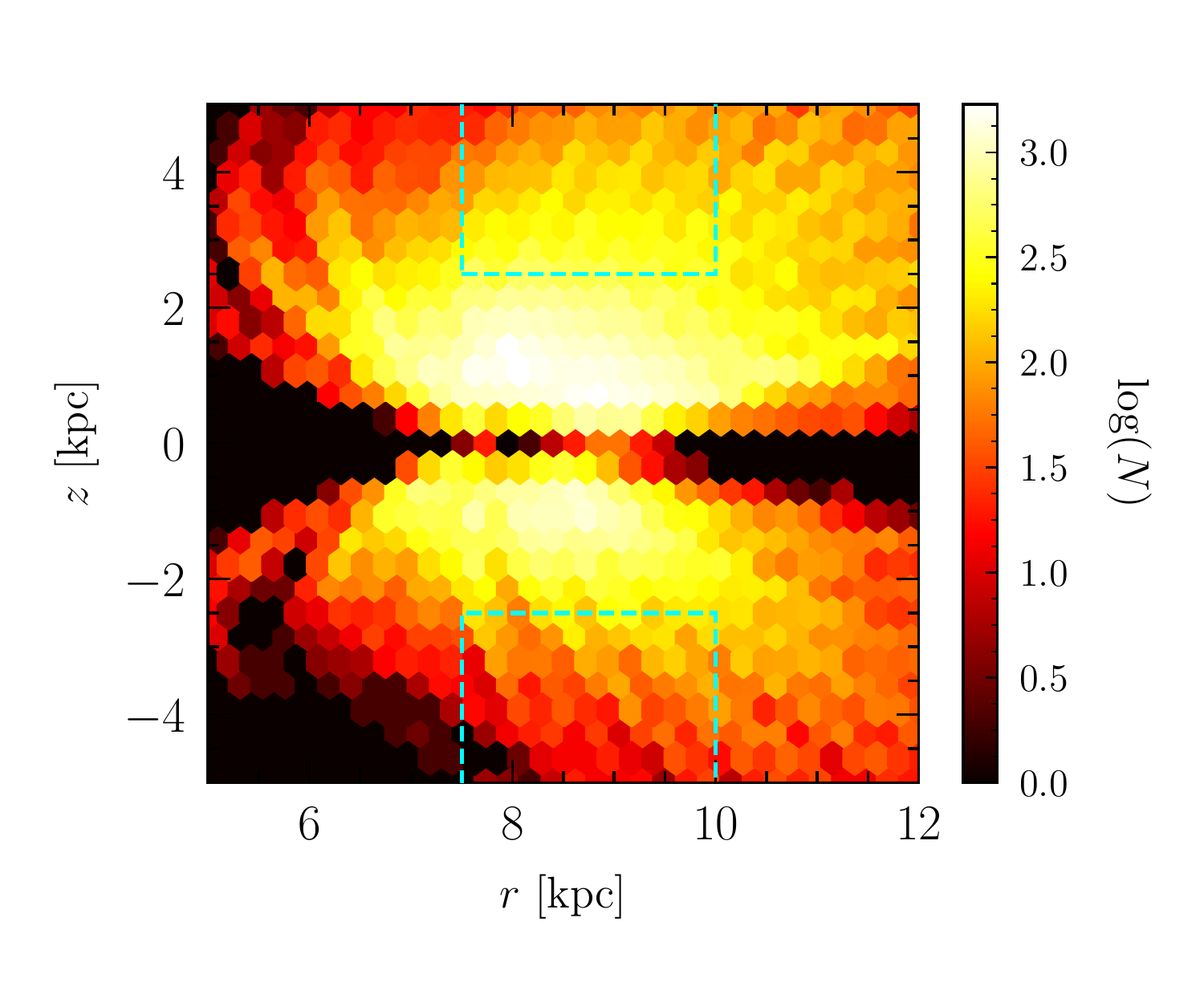}
\caption{The spatial distribution of the SDSS-\emph{Gaia}~DR2 sample in terms of Galactocentric radius, $r$, and vertical distance from the Galactic plane, $z$.  We characterized the disk, halo, and substructure populations in regions within the dashed aqua box, which spans $r\in[7.5, 10.0]$~kpc and $|z| > 2.5$~kpc.}
\label{fig:SDSSspatial}
\end{figure}

\begin{figure*}[t] 
\centering
\includegraphics[width=\textwidth]{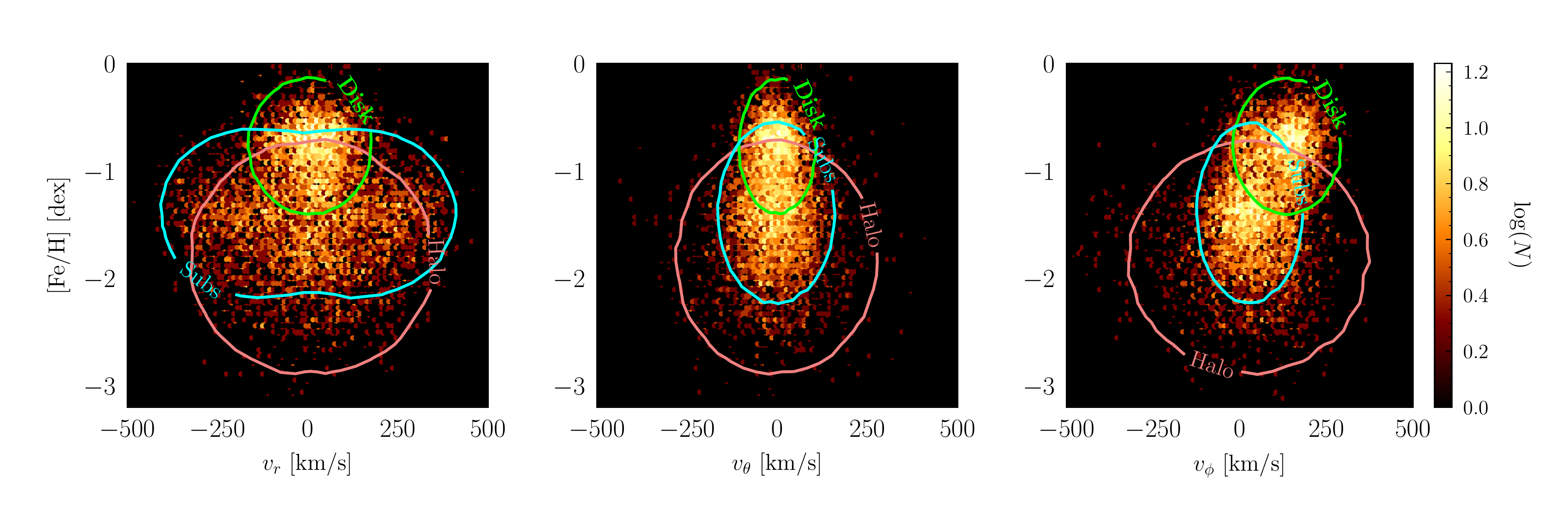}
\caption{Chemo-dynamic distribution of stars in the SDSS-\emph{Gaia}~DR2 sample within $r \in [7.5, 8.5]$~kpc and $|z| > 2.5$~kpc.  The panels show how the distributions vary in iron abundance $\FeH$ and the spherical Galactocentric radial coordinates $v_r$~(left), $v_\theta$~(middle), and $v_\phi$~(right).  The disk population is pronounced at $\FeH \sim -0.8$ and a nearly isotropic halo population is apparent at $\FeH \lesssim -1.8$.  A highly radial population at $\FeH \sim -1.4$ constitutes a large fraction of the sample and is an example of kinematic substructure.  The 95\% contours of the distributions recovered from the likelihood analysis are also shown (see Sec.~\ref{sec:likelihood}); the disk, halo, and substructure best-fits are shown in green, pink, and blue, respectively.}
\label{fig:FeHvpanel}
\end{figure*}

Distances to these predominantly MS stars are calculated
using equations (A2), (A3), and (A7) of
\citet{Ivezic:2008wk}, with distance errors within the region of interest of $\sim 2\%$, or of the order of a $\sim 100$ pc. Recently, the validity of this photometric
parallax calibration was verified by \citet{2018arXiv180510288D} for a
subset of stars with accurate \emph{Gaia} DR2 parallaxes.  We have
also confirmed that the fractional errors on the distance as derived
from the parallaxes are consistently larger than those derived using
the photometric parallax.  The celestial coordinates, heliocentric
distances, proper motions and radial velocities are then used to
calculate stellar velocity components in spherical polar coordinates.
We marginalize over the Local Standard of Rest value assuming that it
is described by a Gaussian with a center at 238 km/s and a dispersion
of 9 km/s \citep[][]{doi:10.1111/j.1365-2966.2012.21631.x}. The
components of the Solar peculiar motion are those presented in
\citet{LSR}. The measured parameters' uncertainties---including
covariances in the proper motion components---are then propagated
using Monte-Carlo sampling to obtain estimates of the uncertainties in
the spherical velocity components \citep[see][for
  details]{2018MNRAS.478..611B}. For the analysis described in the
Sections below, each star's Monte Carlo samples are modeled with a Gaussian
distribution to obtain the full covariance matrix.
  
There are no strong reasons to believe that the SDSS spectroscopic
sample of MS stars used in our analysis suffers from any
appreciable kinematic bias \citep{2018MNRAS.478..611B}. However, as the sample uses a mixture of
SDSS target categories, a moderate metallicity bias towards more
metal-poor stars is expected \citep[see][for details]{Yanny:2009kg}.
We also repeated our benchmark analysis selecting only F/G stars,
which are known to have no biases either in metallicity or
kinematics.  This reduces the sample size substantially, increasing
the uncertainties on the recovered fit parameters.  However, the
overall results---most importantly the fractional contribution of
the stellar components---remain essentially unchanged, within uncertainties.

Fig.~\ref{fig:SDSSspatial} shows the spatial distribution of stellar counts in SDSS-\emph{Gaia} DR2, as a function of Galactocentric radius, $r$, and vertical distance from the Galactic plane, $z$.  

\subsection{Model Motivation}
\label{sec:modelmotivation}

Fig.~\ref{fig:FeHvpanel} shows the distribution of the spherical velocity components as a function of stellar metallicity\footnote{In particular, we use the iron abundance, $\FeH$, which is defined as
\begin{equation}
\FeH = \log_{10}(N_\mathrm{Fe}/ N_\mathrm{H}) - \log_{10}(N_\mathrm{Fe}/ N_\mathrm{H})_\odot \, , \nonumber
\end{equation}
where $N_i$ is the number density of the element $i$.} 
for the SDSS-\emph{Gaia}~DR2 subsample within $r \in [7.5, 8.5]$~kpc and $|z| > 2.5$~kpc.  For Fig.~\ref{fig:FeHvpanel} (and subsequent figures), we sample from the posterior distributions of the mean and covariance matrices to construct the velocity components $v_r, v_\theta, v_\phi$ for the best-fit model. 

Several features in Fig.~\ref{fig:FeHvpanel} are apparent by visual inspection.  First is the disk population, which is centered at $\FeH \sim -0.8$, $v_{r, \theta} \sim 0$~km/s, and $v_\phi \sim 130$~km/s.  Second, is a population with $\FeH \sim -1.4$ with large radial anisotropy.  Third, is a population that extends down to $\FeH \lesssim -1.8$ with nearly isotropic velocities.\footnote{We also note a small cluster of stars at $\FeH \sim -1.5$ and $v_\phi < -200$~km/s.  These stars may or may not be part of a distinct population; because they constitute only 0.5\% of the total sample, we will not focus on them in this work.  They are primarily flagged as `halo' stars in the likelihood study.}   

In this work, we will refer to the second population as `substructure' and the third population as the `halo.'  We are envisioning that both originate from the disruption of accreted satellites in the Milky Way.  What we refer to as the `halo' is intended to encapsulate the tidal debris from the oldest mergers, which will typically be the most metal-poor and fully well-mixed in phase space.   What we call `substructure' constitutes tidal debris that is not fully phase mixed; such a component may exhibit interesting features in spatial and/or velocity coordinates,  such as streams or debris flow.  The prefix `sub-' suggests that this population is less dominant than the halo population; we adopt this terminology as it is standard in the DM literature, but make no assumptions on its relative dominance in our study.

Evidence has been building for a multi-component inner stellar halo that is dominated by the tidal debris of one massive merger~\citep{2015MNRAS.448L..77D, 2015ApJ...798L..12F,2018MNRAS.477.1472B, 2018arXiv180606038H, 2018ApJ...856L..26M}.  The large radial anisotropy of the stars with $\FeH \gtrsim -1.7$  in Fig.~\ref{fig:FeHvpanel} was first identified using the SDSS-\emph{Gaia} DR1 sample~\citep{2018MNRAS.478..611B}.\footnote{Note that we define $\phi$ and $\theta$ as the azimuthal and polar directions, respectively.  This is the opposite of the convention used in~\cite{2018MNRAS.478..611B}.} This work noted that the `sausage'--like feature in the data appears to be non-Gaussian and estimated its contribution to be $\sim66\%$ of the non-disk population over the full SDSS footprint.  They found that the radial anisotropy of the sample drops markedly at $\FeH \lesssim -1.7$, suggesting that a separate isotropic and metal-poor population is also present.  It is unlikely that the radial and  isotropic populations originated in the Milky Way as their iron abundances are in-line with those observed in Milky Way dwarf spheroidal galaxies~\citep{2004AJ....128.1177V} and their velocities are distinct from disk stars.

Recent work using local MS stars from SDSS-\emph{Gaia}~DR2 as well as a separate sample of more distant Blue Horizontal Branch stars demonstrated that the orbits of the most highly eccentric stars share a common apocenter at $r \sim 20$~kpc~\citep{2018arXiv180510288D}.  This radius is coincident with the observed break in the Milky Way's stellar density distribution~\citep{Deason:2012fc}, suggesting that the radial stellar population is the tidal debris of a recent and large merger that dominates the inner halo.  Simulations have shown that the density of stars from a particular merger may exhibit a break at the average apocenter of the stars~\citep{Deason:2012fc}.  

If the radial substructure is indeed associated with a large merger, one might expect that globular clusters were also stripped from the satellite progenitor as it was disrupted.  Indeed, a number of globular clusters on highly radial orbits were recently identified that may be associated with the large merger(s) causing the sausage-like feature in the SDSS-\emph{Gaia} data~\citep{Myeong:2018kfh}.  The number of these clusters suggests a total progenitor mass of $\sim 10^{10}$~M$_\odot$; their tracks in age-metallicity space bound the maximum infall redshift less than $\sim 3$.   

These new results have direct implications for the local DM distribution.  Previous work demonstrated that the virialized DM component is traced by the most metal-poor stars in the Milky Way---this corresponds to the population that we refer to as the halo here~\citep{eris_paper,Necib:2018}. The virialized DM is well-tracked by stars with metallicities $\FeH \lesssim -3$ in the \textsc{Eris} simulation~\citep{eris_paper}, and for $\FeH \lesssim -2$ in the~\textsc{Fire} simulation~\citep{Necib:2018}.  The relevant metallicity range depends on the time when the earliest mergers occurred, which depends on a galaxy's detailed merger history.  Additionally, the kinematic substructure observed in the data is highly reminiscent of debris flow~\citep{Lisanti:2011as, Kuhlen:2012fz}.  Indeed, a study of the stellar halo in \texttt{Via Lactea} (where star particles were painted onto the most bound DM particles in subhalos) found precisely the same kind of radial substructure becoming apparent in the SDSS-\emph{Gaia} data, and that the kinematics of the accreted stars correlate with that of the DM debris~\citep{Lisanti:2014dva}.  The stellar-DM correlation for debris flow has since been verified using the \textsc{Fire} simulations~\citep{Necib:2018}. Therefore, if we want to infer the kinematic properties of the local DM, we will need to model the velocities of the halo and substructure populations.

\begin{table}[t]
\begin{center}
\begin{tabular}{C{1.6cm}C{0.8cm}C{1.5cm}C{1.5cm}C{1.5cm}}
\Xhline{3\arrayrulewidth}
\renewcommand{\arraystretch}{1}
Parameter & Type&\multicolumn{3}{c}{Priors} \\
&& Disk & Halo & Sub \\ 
\hline
$\mu_r$& linear&  $[-70,70]$ & $[-70,70]$  & $[0, 250]$\\
$\mu_\theta$& linear&  $[-70,70]$ & $[-70,70]$ & $[-70,70]$\\
$\mu_{\phi}$& linear& $[0,300]$ & $[-70, 70]$& $[-70,70]$ \\
$\sigma_{r, \theta, \phi} $& linear &$ [0,200]$ & $[0, 200]$& $[0, 200]$ \\
$\rho_{{r \theta}, {r \phi}, {\theta \phi}}$& linear & $[-1,1] $ & $[-1,1]$ & $[-1,1]$ \\
$\mu_{\scriptscriptstyle \FeH}$ & linear & $[-1.5, 0.5]$  & $[-3, -1]$ & $[-3, -1]$\\
$\sigma_{\scriptscriptstyle{\FeH}}$ & linear & $ [0, 2]$ & $[0, 2]$& $[0,2]$\\ 
$Q$& linear & --- & $[0, 1]$ & $[0,1]$  \\
\Xhline{3\arrayrulewidth}
\end{tabular}
\caption{\label{tab:priors} Parameters and associated prior types/ranges for the disk, halo, and substructure populations.}
\end{center}
\end{table}

\vspace{0.5cm}
\subsection{Likelihood Procedure}
\label{sec:likelihood}

To isolate the accreted stellar population, we can place a hard upper cut-off on the metallicity of the sample.  The downside to this conservative approach is that it ignores the high-metallicity tail of the accreted stellar distribution that overlaps with  disk stars.  For this reason, we use a mixture model analysis to statistically identify the individual populations of accreted stars over the full metallicity range of the sample.  
 
Each star, labeled by the index $i$, is associated with a set of observable quantities such as its velocity and metallicity, $O_i = \left(\mathbf{v}_i, \FeH_i\right)$, as well as the variance for each.  We assign each star a flag $j = d, h, s$ that designates whether it belongs to the disk, halo, or substructure population, respectively.  The likelihood of observing $O_i$ for a disk star is 
\begin{eqnarray}
p_d \,(O_i \, |\, \theta ) = \mathcal{N} \left( \mathbf{v}_i \,| \, \boldsymbol{\mu}^d, \boldsymbol{\Sigma}^d_i \right) 
 \mathcal{N} \left( \FeH_i \,| \, \mu^d_{\scriptscriptstyle \FeH}, \sigma^d_{{\scriptscriptstyle{\FeH}},i} \right), \nonumber\\
&&  
\label{eq:disklikelhood}
\end{eqnarray}
where $\theta$ is the set of free parameters and $\mathcal{N}$ denotes the normal distribution.  The set $\theta$ includes the velocity distribution mean  $\boldsymbol{\mu}^d$ and covariance matrix $\boldsymbol{\Sigma}^d_i$, as well as the metallicity distribution mean $\mu^d_{\scriptscriptstyle \FeH}$  and dispersion $\sigma^d_{{\scriptscriptstyle{\FeH}},i}$.  The covariance matrix depends on the individual velocity dispersions $\sigma_{r, \theta, \phi}$ as well as the correlation coefficients $\rho_{r\theta}, \rho_{r\phi}, \rho_{\theta \phi}$.  Note that the velocity covariance matrix and the metallicity dispersion vary between stars because the observed covariance depends on the true value and the measurement error---specifically, $\boldsymbol{\Sigma}_\mathrm{obs} = \boldsymbol{\Sigma}_\mathrm{true} + \boldsymbol{\Sigma}_\mathrm{err}$.  There are eleven parameters associated with this model.  The likelihood for a halo star is also given by~\eqref{eq:disklikelhood}, except with $d\rightarrow h$, and thus comes with an additional eleven parameters. 

\begin{figure*}[t] 
\centering
\includegraphics[width=0.75\textwidth]{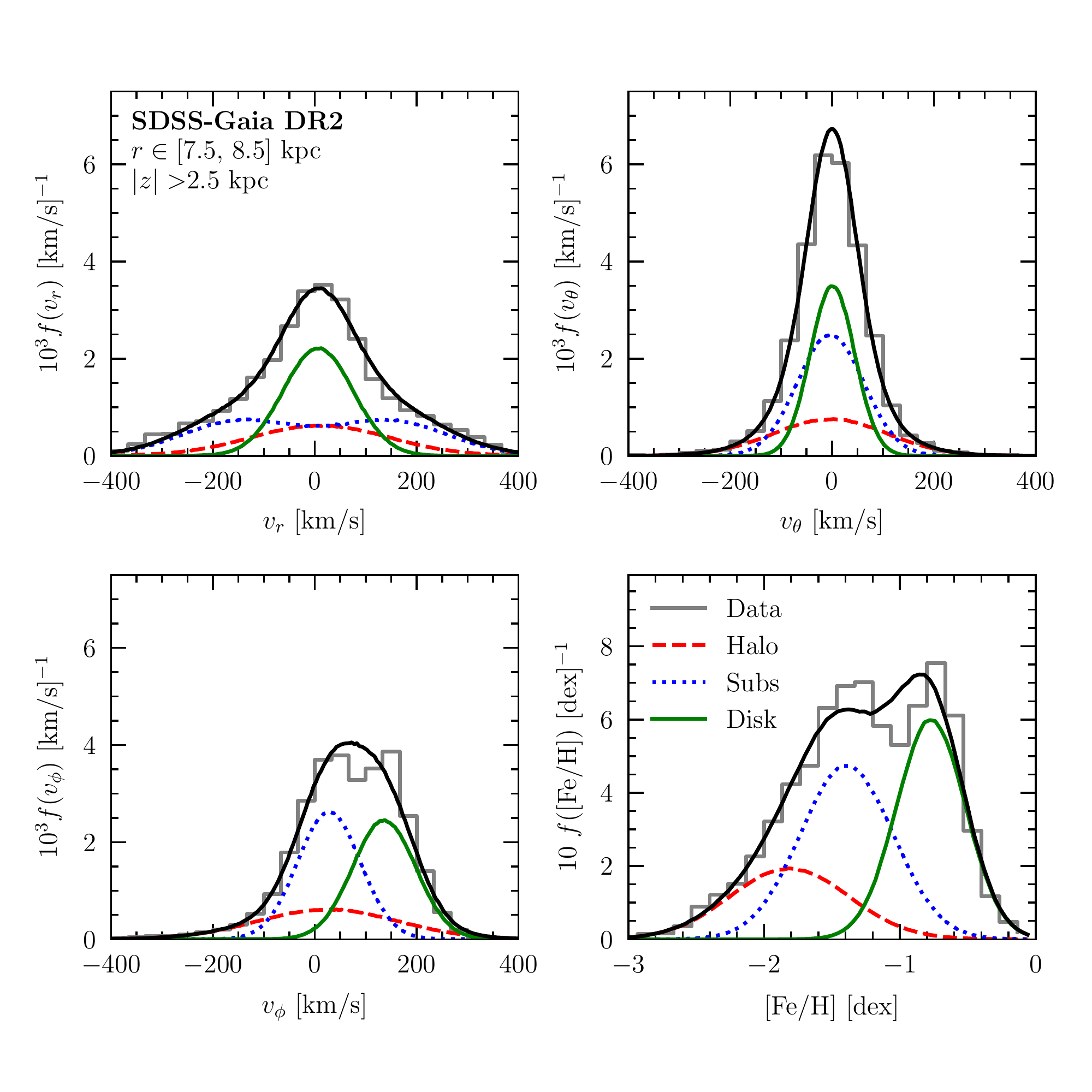}
\caption{Best-fit distributions of the spherical Galactocentric velocity components and metallicity for the SDSS-\emph{Gaia}~DR2 sample in the region within $r \in [7.5, 8.5]$~kpc and $|z| > 2.5$~kpc (clockwise from top left: $v_r$, $v_\theta$, $\FeH$, and $v_\phi$).  These distributions are found from sampling the posteriors of the model parameters.  In each panel, the disk, halo, and substructure distributions are shown as solid green, dashed red, and dotted blue lines, respectively.  The data is represented by gray histograms.  The corresponding corner plots are shown in the Appendix as Figs. \ref{fig:corner_halo}, \ref{fig:corner_disk}, and \ref{fig:corner_lobes}.
}
\label{fig:Vposteriors}
\end{figure*}

\begin{figure*}[t] 
\centering
\includegraphics[width=0.75\textwidth]{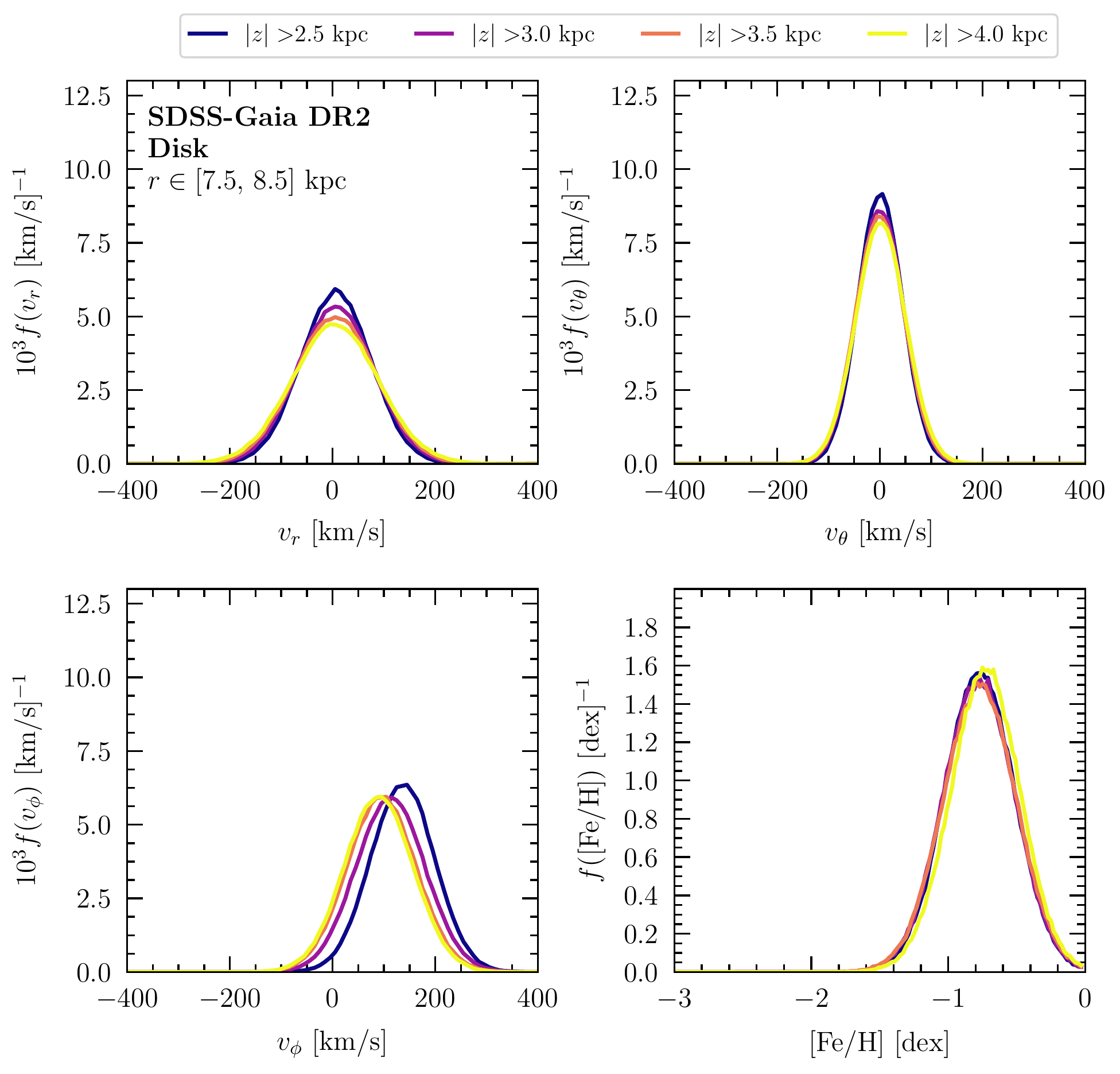}
\caption{Best-fit distributions for the disk population in the region with $r \in [7.5, 8.5]$~kpc (clockwise from top left: $v_r$, $v_\theta$, $\FeH$, and $v_\phi$).  We vary the distance from the mid-plane from $|z| > 2.5$ to $4.0$~kpc. }
\label{fig:zcutDisk}
\end{figure*}

\begin{figure*}[t] 
\centering
\includegraphics[width=0.75\textwidth]{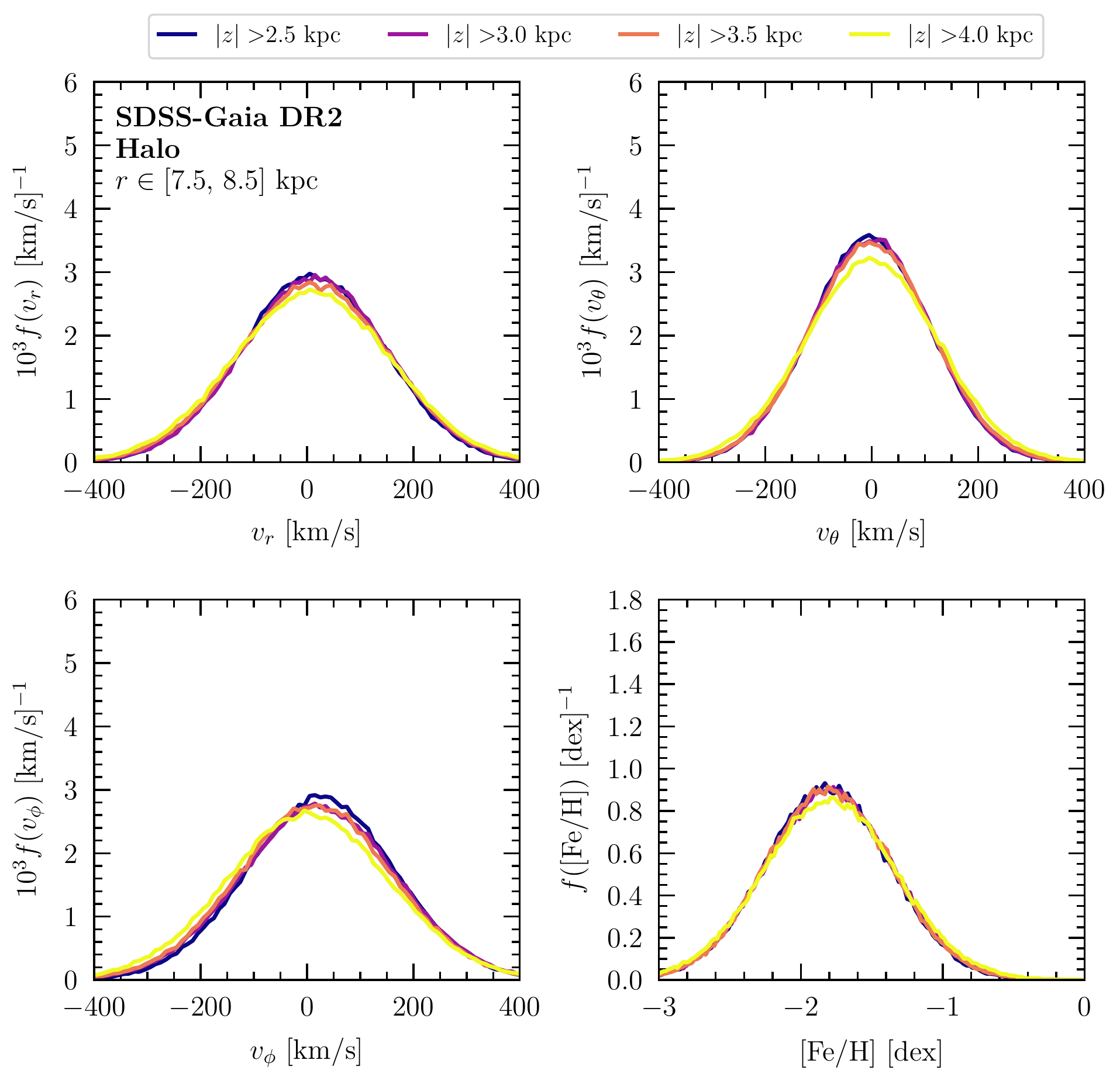}
\caption{The same as Fig.~\ref{fig:zcutDisk}, except for the halo population.  The chemo-dynamic properties of the halo are invariant as one moves away from the Galactic mid-plane.}
\label{fig:zcutHalo}
\end{figure*}

\begin{figure*}[t] 
\centering
\includegraphics[width=0.75\textwidth]{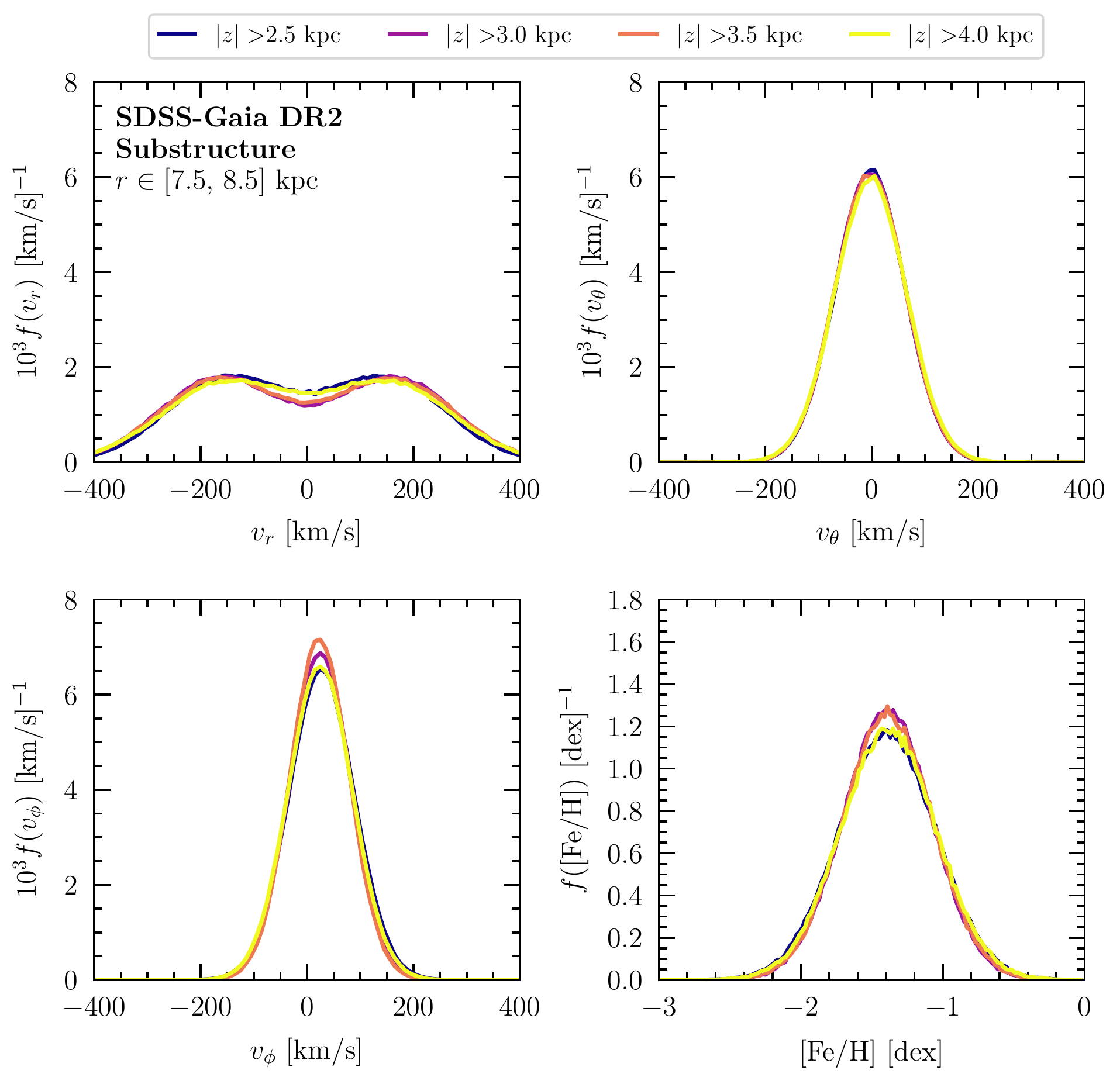}
\caption{The same as Fig.~\ref{fig:zcutDisk}, except for the substructure population.  The radial distribution has  characteristic lobes at $\pm 148$~km/s that are likely related to tidal debris that is stripped as a merging satellite moves towards/away from the Galactic Center on its orbit.  The distributions remain constant over the entire $z$-range.}
\label{fig:zcutSubs}
\end{figure*}

Modeling the substructure population is more challenging as initial evidence suggests that its radial velocities are non-Gaussian~\citep{2018MNRAS.478..611B}.  Therefore, we assume that the velocities are a sum of two multivariate normal distributions with equivalent parameters, except for equal and opposite mean in $v_r$:
\begin{eqnarray}
p_s\, (O_i \, |\, \theta ) =  \frac{1}{2} \, [ \, \mathcal{N} &&\left( \mathbf{v}_i | \boldsymbol{\mu}^{\tilde{s}}, \boldsymbol{\Sigma}^s_i \right) +  \mathcal{N} \left( \mathbf{v}_i | \boldsymbol{\mu}^{s},  \boldsymbol{\Sigma}^s_i \right)  \,] \,  \nonumber \\
&& \times \,  \mathcal{N} \left( \FeH_i \,| \, \mu^s_{\scriptscriptstyle \FeH}, \sigma^s_{{\scriptscriptstyle{\FeH}},i} \right), 
\label{eq:sublikelihood}
\end{eqnarray}
where $\boldsymbol{\mu}^{\tilde{s}} = (-\mu_r, \mu_\theta, \mu_\phi)^{s}$.  This model can vary over unimodal and bimodal distributions in $v_r$.  For example, when $\mu_r^s \rightarrow 0$,~\eqref{eq:sublikelihood} approaches a single Gaussian distribution peaked at zero.  In the limit where $\mu_r^s \gg \sigma_r^s$ and $\mu_r^s \neq 0$, then the radial lobes are very pronounced.  If, in contrast, $\mu_r^s \ll \sigma_r^s$, then the overlap between the two lobes increases and the radial velocity distribution is more box-like.  We assume that the radial lobes, if present, are symmetric about $v_r = 0$~km/s, as this would be expected if the tidal debris originates from a satellite as it moves towards ($v_r < 0$) and then away from ($v_r > 0$) the Galactic Center.\footnote{For a visual example of the bimodal distribution, we point the reader to the top left panel of Fig.~\ref{fig:zcutSubs}, which we will discuss in more detail in the following section.}

The likelihood for the complete set of $N$ stars is
\begin{eqnarray}
p\left( \{ O_i \} \,|\, \theta \right) &=& 
 \prod_{i=1}^N \, \sum_{j=d,h,s} Q_j \,p_j \left( O_i \, |  \,\theta \right), 
   \label{eq:totallikelihood}
\end{eqnarray}
where the brackets around the $O_i$ indicate the full list of $N$ values.  $Q_j$ is the probability that the star belongs to the $j^\mathrm{th}$ population; these represent two additional parameters in the model, as $Q_h = 1- Q_d- Q_s$.   

We use \texttt{emcee}~\citep{2013PASP..125..306F} to find the posterior distributions of all 35 free parameters.  In particular, we use 250 walkers, with 5000 steps, and a burn-in period of 10000 steps.  The priors for the separate parameters are provided in Table~\ref{tab:priors}.  We perform the mixture analysis in separate regions within the dashed aqua box of Fig.~\ref{fig:SDSSspatial}, which spans from $r \in [7.5, 10.0]$~kpc and $|z| > 2.5$~kpc. We find that the fit is well-behaved in this radial span, as gauged primarily by its ability to reproduce the expected properties of the baryonic disk.  Below $|z| \sim 2.5$~kpc, we find a persistent systematic bias in the fitting procedure that results from modeling the azimuthal disk velocities as a single Gaussian~\citep{2012MNRAS.419.1546S}, so we do not present those results here.

\section{The Stellar Distribution}
\label{sec:stellarhalo}
The 95\% contours in metallicity--velocity space that are recovered from the analysis are overlaid on the separate panels of Fig.~\ref{fig:FeHvpanel}.  These results apply specifically to the SDSS-\emph{Gaia}~DR2 sample in the region where $r \in [7.5, 8.5]$~kpc and $|z| > 2.5$~kpc.  Clearly, the best-fit distributions do an excellent job of picking out the population clusters that we identified by eye at the offset.  Note that the priors are, for the most part, uninformative.  They make no assumption about the relative metallicities of the halo and substructure populations, or the means and dispersions of their velocity distributions.  The choice of priors gives the analysis enough freedom to explore both unimodal and bimodal distributions for the radial velocity of the substructure.

Fig.~\ref{fig:Vposteriors} shows the one-dimensional $v_r, v_\theta, v_\phi,$ and $\FeH$ best-fit distributions for the same region.  The disk, halo, and substructure distributions are indicated by the solid green, dashed red, and dotted blue lines, respectively, while the data is represented by the gray histograms.  In the Appendix, we also provide the residuals between the model and data as \Fig{fig:residual}, as well as the corresponding corner plots as Figs. \ref{fig:corner_halo}, \ref{fig:corner_disk}, and \ref{fig:corner_lobes}.  One of the most surprising results is the degree to which the halo distribution is subdominant relative to the substructure.  To better understand this, let us review in some detail the behavior of each population separately, and characterize its  evolution as we vary the lower bound on $|z|$ in the range from 2.5--4.0~kpc.   

From Fig.~\ref{fig:Vposteriors}, it is clear that the best-fit disk population has a median metallicity of $\FeH = -0.77$\footnote{The errors on the best-fit metallicity means and dispersions quoted here are all on the order $\pm 0.01$~dex.} and $\mu_\phi = 136.3^{+1.9}_{-1.8}$~km/s, consistent with the data. Fig.~\ref{fig:zcutDisk} demonstrates how the disk's velocity and metallicity distribution vary away from the mid-plane.  The $\FeH$ and $v_\theta$ distributions remain essentially constant as one moves from $|z| > 2.5$ kpc to $> 4$~kpc.  The $v_r$ distribution broadens slightly and the median $v_\phi$ shifts to lower values, as expected from asymmetric drift~\citep{Bond:2009mh}

Fig.~\ref{fig:zcutHalo} shows the corresponding distributions for the halo, which remain constant over the full $z$-range explored here.  This population is clearly very metal-poor with a median $\FeH = -1.82$. Its velocity distribution is nearly isotropic as $\sigma_r = 136.1^{+3.6}_{-3.6}$~km/s, $\sigma_\theta = 112.5^{+4.1}_{-3.8}$~km/s, and $\sigma_\phi = 139.1^{+5.5}_{-5.2} $~km/s.  The  radial and azimuthal means are non-zero, with $\mu_r = 10.0_{+4.9}^{-4.6}$ km/s and $\mu_\phi = 24.9_{+4.6}^{-5.6}$ km/s.  All three correlation coefficients are small: $(\rho_{r \theta}, \rho_{r \phi}, \rho_{\theta \phi}) = (-0.03^{+0.03}_{-0.03}, -0.08^{+0.03}_{-0.03}, 0.06^{+0.01}_{-0.02})$.

The halo is subdominant to the substructure, which is distinctive in both chemical abundance and kinematics.  As shown in Fig.~\ref{fig:zcutSubs}, the median metallicity of the substructure remains constant at $\FeH = -1.39$ over all $z$-values.  This population is more metal-rich, on average, than the halo, but more metal-poor than the disk.  The radial velocity lobes are centered at $\mu_r = \pm 147.6^{+7.2}_{-6.4}$~km/s with $\sigma_r = 113.6^{+3.1}_{-3.0}$~km/s.  There is no evidence for rotation in the polar direction ($\mu_\theta = -2.8_{-1.6}^{+1.5}$ km/s and $\sigma_\theta = 65.2_{-1.2}^{+1.1}$~km/s), however there is a larger offset in the azimuthal direction, with $\mu_\phi = 27.9_{-2.9}^{+2.8}$ km/s and $\sigma_\phi = 61.9_{-2.9}^{+2.6}$~km/s. The correlations $\rho_{r \theta}$ and $\rho_{\theta \phi}$ are consistent with zero, while $\rho_{r \theta} = 0.18^{+0.03}_{-0.03}$. 

\begin{figure*}[t] 
\centering
\includegraphics[width=\textwidth]{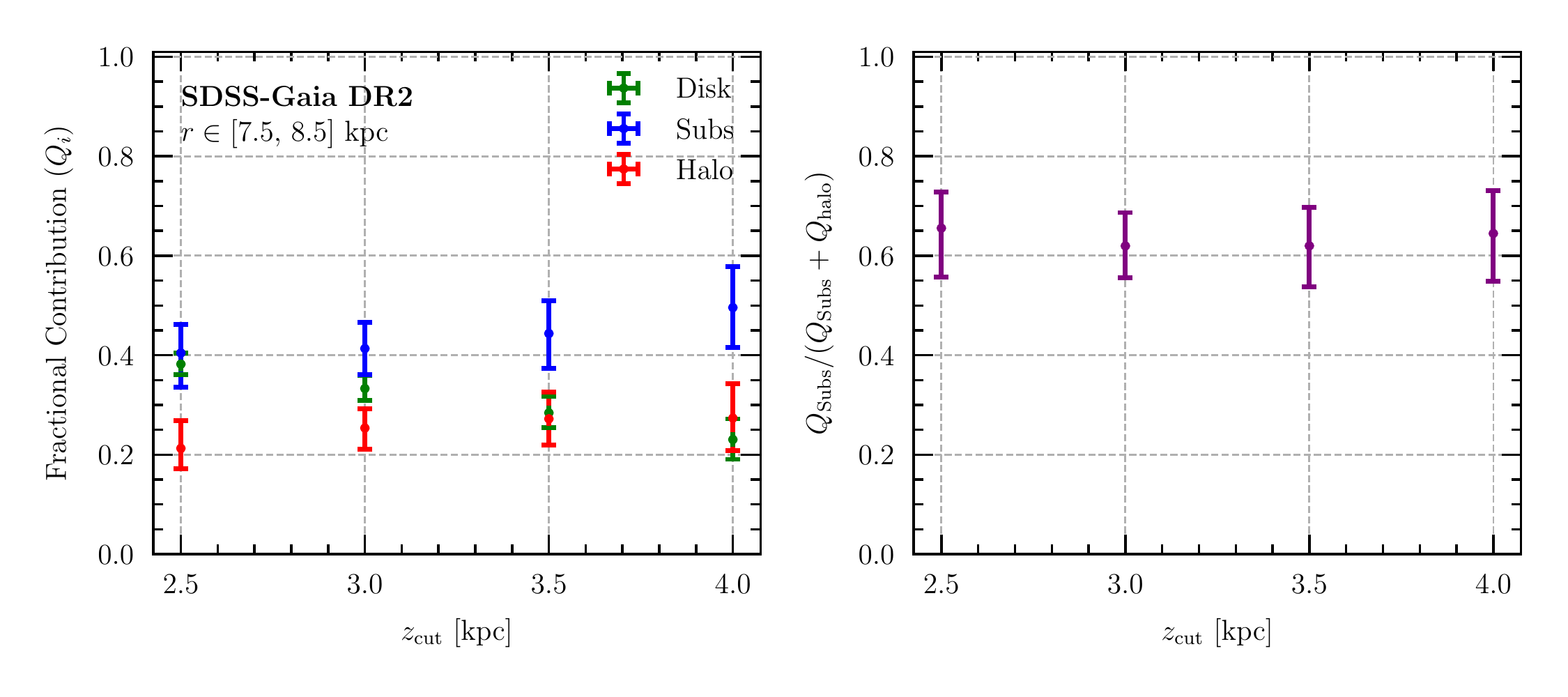}
\caption{(Left) The fractional contribution of the disk (green), halo (red), and substructure (blue) populations in the dataset,  for the region $r\in[7.5, 8.5]$~kpc and $|z|> z_\mathrm{cut}$.  The 2.1, 50, and 97.9$^\text{th}$ percentiles are shown here. (Right)  The fraction of the substructure relative to all non-disk stars in the dataset. }
\label{fig:fractions}
\end{figure*}

Fig.~\ref{fig:fractions} shows the fractional contribution of the disk, halo, and substructure stars in the dataset.  We see that the disk contribution reduces from 40\% at $|z| > 2.5$~kpc, down to 25\% at $|z| > 4$~kpc.  The halo contribution increases mildly in this range, as might be expected.  However, the relative fraction of the substructure to the non-disk stellar population (right panel) is constant at $\sim 60$\%.  We remind the reader that these fractional contributions pertain only to the dataset, and that metallicity biases can potentially affect the extrapolation to the Galaxy.  However, the fact that the results are unchanged (within uncertainties) when the analysis is repeated on the subset of F/G stars, which exhibit minimal bias, gives us confidence in the results presented here. 

We made the corresponding versions of Fig.~\ref{fig:zcutDisk}--\ref{fig:fractions} for  $|z| > 2.5$~kpc and varying $r$ in five equally sized bins from 7.5--10.0~kpc.  These figures are provided in the Appendix.  We observe no significant change in the velocity and metallicity distribution of the halo in this range.  The substructure distributions are also stable, except that the radial lobes are further apart and more pronounced  closer to the center of the Galaxy, as shown in \Fig{fig:Subs_rgc3kpc}.  The mean of the radial distribution is at $\mu_r = 140.5^{+8.4}_{-7.7}$ km/s with $\sigma_r = 114.5^{+4.1}_{-3.9}$~km/s for $r \in [7.5, 8.0]$~kpc, while it drops to $\mu_r = 115.0^{+3.1}_{-3.1}$~km/s with $\sigma_r = 104.2^{+2.7}_{-2.6}$~km/s for $r \in[9.5, 10]$~kpc.  This may be related to features of the orbit that change with $r$.

The trends we observe are consistent with the interpretation that the substructure originates from the merger of a fairly massive satellite on a highly radial orbit~\citep{2018MNRAS.477.1472B,2018arXiv180510288D,Myeong:2018kfh,2018ApJ...856L..26M}. We distinguish two individual lobes in the radial velocity distribution, consistent with tidal debris that is preferentially stripped as the satellite moves towards/away from the Galactic Center.  The small, but non-zero, azimuthal rotation may also be linked with the properties of the orbit.  We note that our analysis cannot distinguish between one or more mergers.  The latter situation seems unlikely as the metallicity of the substructure remains constant over the entire spatial range probed, which suggests a single progenitor.  If multiple mergers were at cause, the satellites would have to have similar masses and orbital properties, which seems fine-tuned. 

It is challenging to directly compare our best-fit values to previous studies of the stellar halo, because we break down the sample into a Gaussian and non-Gaussian component at the likelihood level.  As a result, the halo velocity distribution published in  other works would be the weighted sum of our halo and substructure populations.  However, the general trends we observe are roughly consistent with previous results.  For example, previous studies found that halo stars with intermediate metallicities ($\FeH\sim -1.4$) are radially anisotropic~\citep{Bond:2009mh,Smith:2009kr,2018MNRAS.478..611B}, while stars on the more metal-poor end of the spectrum become more isotropic~\citep{Carollo:2007xh, Carollo:2009dz, Herzog-Arbeitman:2017zbm, 2018MNRAS.478..611B}.

\section{The Dark Matter Distribution}
\label{sec:darkmatter}

We now discuss how our results concerning the accreted stellar population translate to the DM distribution.  In particular, we will argue that the presence of an anisotropic population of stars suggests that a component of the DM halo may have also originated from a more recent merger, and may consequently not have reached steady state.  Sec.~\ref{sec: twocomponent} outlines a simple model framework for this two-component DM scenario, and Sec.~\ref{sec: directdetection} describes the consequences for direct detection experiments.

\subsection{Two-Component Dark Matter Model}
\label{sec: twocomponent}

Let us assume that the DM in the region of study can be divided into two populations.  For the moment, we remain agnostic to the origin of these two populations, but we will return to this shortly.  In this scenario, the total DM phase-space distribution satisfies 
\begin{equation}
f(v) = \xi_1 \, f_1(v) + \xi_2 \,  f_2(v) \, ,
\label{eq: twocomponent}
\end{equation}
where $f_i$ is the normalized velocity distribution and $\xi_i$ is the relative fraction of the $i^\text{th}$ population.  Note that $\xi_1 + \xi_2 = 1$.  Starting from the distribution function in \Eq{eq: twocomponent}, it is possible to derive the Jeans equations in spherical coordinates.  To simplify the derivation and highlight the most important conceptual points, we make several assumptions about these two populations.  

First, we assume that Population~1 is isotropic, spherically symmetric, and that its velocity components are uncorrelated.  Most importantly, we take it to be in steady state, which allows us to ignore partial derivatives of $f_1(v)$ with respect to time.  When we state that a population is in equilibrium, we mean specifically that it is in steady state with  $\partial f_i/ \partial t = 0$.

As a point of contrast, we will assume that Population~2 has not reached steady state---or, at the very least, that we cannot confirm whether it has.  Additionally, we will assume that its spatial density is spherically symmetric, that its velocity components are uncorrelated and have  vanishing mean at present-day, and that the mean velocities and dispersions are spatially invariant in the region of interest.

The radial Jeans equation for this two-component model is
\begin{widetext}
\begin{equation}
 \xi_2 \, \nu_2  \, \frac{\partial \mu_{r,2}}{\partial t}  +  \left[ \xi_1\,\sigma_{r,1}^2 \frac{\partial \nu_1}{\partial r} + \xi_2 \, \sigma_{r,2}^2  \, \frac{\partial \nu_2}{\partial r} \right]+ \frac{\xi_2 \, \nu_2}{r} \left[ 2\sigma_{r,2}^2 - \sigma_{\phi,2}^2 - \sigma_{\theta, 2}^2 \right] =-\left(\xi_1 \nu_1 + \xi_2 \nu_2 \right) \frac{\partial \Phi}{\partial r} \, ,
\label{eq: radialJeans}
\end{equation}
\end{widetext}
where $\nu_i$ is the number density, $\mu_{r,i}$ is the radial velocity mean, and $\sigma_{r,i}, \sigma_{\theta, i}, \sigma_{\phi, i}$ are the velocity dispersions for the $i^\text{th}$ population and $\Phi$ is the gravitational potential. 

Using \Eq{eq: radialJeans}, one can recover the classical argument laid out in \cite{Drukier:1986tm} that motivates a Maxwell-Boltzmann velocity distribution for DM.  The observation of a flat rotation curve near the Solar position suggests a logarithmic potential for the Milky Way halo of the form $\Phi(r) = v_c^2 \ln(r)+\text{ constant}$, where $v_c$ is the circular velocity.  
If we assume that all of the DM is in steady state ($\xi_1 = 1$, $\xi_2 = 0$) and that its number density is described as a falling power law, $\nu_1 \propto r^{-b}$, then \Eq{eq: radialJeans} predicts a tight link between the value of the power-law index $b$, the potential $\Phi$, and the phase-space distribution.  In this limit, the radial Jeans equation becomes
\begin{equation}
\frac{\sigma_{r,1}^2}{\nu_1} \frac{\partial \nu_1}{\partial r} = - \frac{v_c^2}{r}  \quad \longrightarrow \quad b = \frac{v_c^2}{\sigma_{r,1}^2} \,.  
\label{eq:simplecase}
\end{equation}
Taking the local circular velocity to be  $v_c \sim 235$~km/s and a dispersion of $\sigma_{r,1} \sim 160$~km/s yields a power-law slope of $b \sim 2$.  A density distribution of the form $\nu_1 \sim r^{-2}$ is isothermal and one can use Poisson's equation to show that the associated velocity distribution is the Maxwell Boltzmann.  

However, notice from \Eq{eq: radialJeans} that that the simple derivation of the isothermal model breaks down in the presence of a second DM population.  In particular, the tight predictive link between the potential $\Phi$ and the number density $\nu_1$ no longer holds when $\xi_2 \neq 0$.  This is true regardless of whether the second DM population is in steady state or not.  However, the challenge is exacerbated for the latter case, as we then need to quantify $\partial \mu_{r,2}/\partial t$.

This two-component model for the DM distribution\footnote{Specifically, the dark matter halo in the region of study within Galactocentric radii of 7.5--10~kpc and $|z| > 2.5$~kpc.} is motivated in light of recent observations of the nearby accreted stellar distributions.  As already discussed, the observation of a metal-poor and isotropic stellar halo is likely associated with tidal debris from the oldest luminous mergers that built up the Milky Way, while the anisotropic component at intermediate metallicities is due to tidal debris from a more recent merger.  As DM would have also been stripped from these accreted satellites, it is reasonable to assume that it should also be separated into at least two populations as well.  It is of course possible that the DM can be divided into more than two populations.  This would be relevant if a significant fraction of the DM originated from non-luminous satellites or diffuse accretion, neither of which should be correlated with stellar distributions.

For now, we solely focus on the subset of DM from \emph{luminous} mergers, as that is where the results of this work have the most consequence.  In this case, the recent observations from \emph{Gaia} suggest that we should be moving to a two-component model for the DM, as per \Eq{eq: twocomponent}.  Population~1 would correspond to the DM from the oldest mergers (the `halo' component) and Population~2 would correspond to the DM from the recent merger (the `substructure' component).

Numerical simulations have demonstrated that DM accreted from the oldest luminous mergers are well-traced by metal-poor stars---this corresponds to the halo population~\citep{eris_paper, Necib:2018}.  Additionally, DM in debris flow is well-traced by intermediate metallicity stars~\citep{Lisanti:2014dva, Necib:2018}.  This suggests that one can use the stellar velocity distribution for the halo and substructure populations, recovered in Sec.~\ref{sec:stellarhalo}, to model the individual DM populations.

As the halo population is the most metal-poor and is also isotropic, we expect that it consists of tidal debris from the earliest mergers that have reached stead state today.  The metallicity range and velocity profiles that we recover share similar features to the distribution of stars accreted at redshifts $z > 3$ in a study of two separate Milky-Way--like halos in the suite of \textsc{Fire} hydrodynamic simulations~\citep{Necib:2018}.  We believe that this population of tidal material is accreted while the proto-galaxy is still forming, and is fully relaxed today. 

The substructure population arises from a more recent merger, with estimated accretion redshift between $z \sim 1$--3~\citep{2018MNRAS.477.1472B,Myeong:2018kfh}.  \cite{Necib:2018} showed that the tidal debris from satellites accreted at $z\lesssim3$ leave distinctive features in velocity space that stand out from the behavior of older tidal debris, which is fully relaxed. Further study is required to determine whether this new substructure has reached steady state. 

\subsection{Experimental Implications}
\label{sec: directdetection}

\begin{figure*}[t] 
\centering
\includegraphics[width=0.45\textwidth]{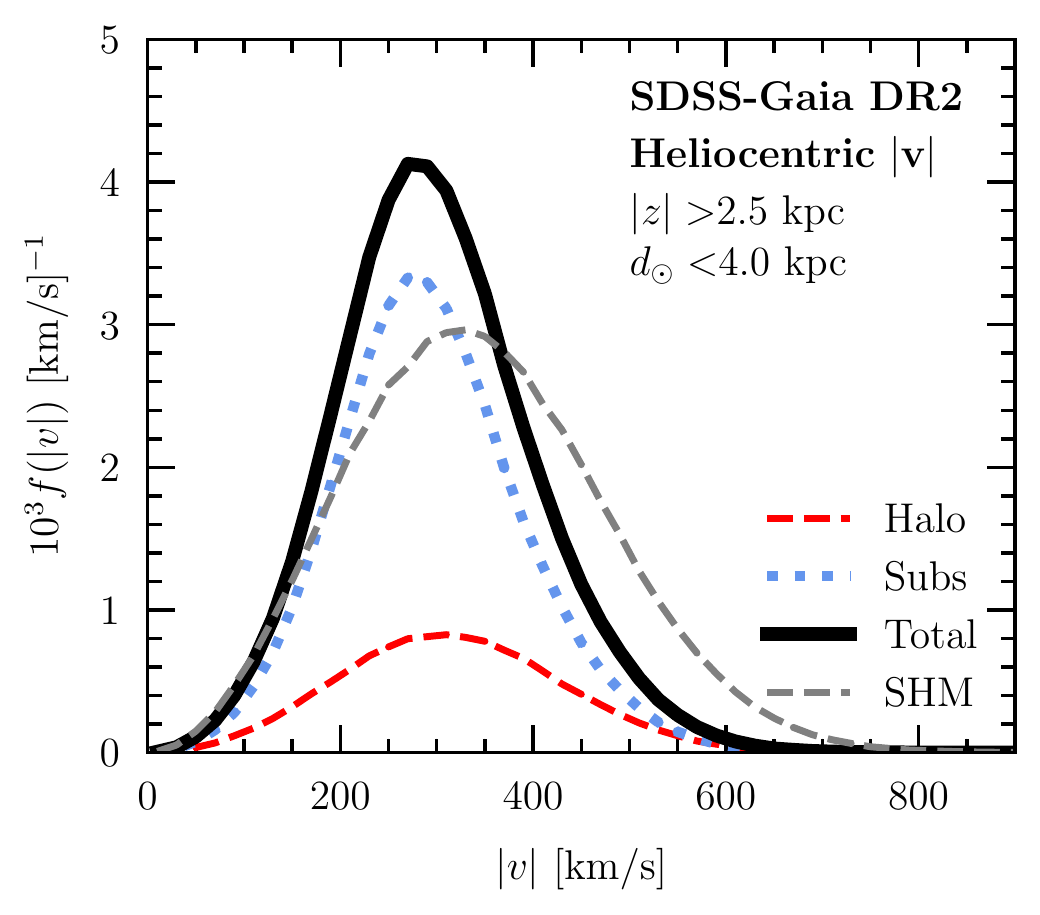}
\qquad
\includegraphics[width=0.49\textwidth]{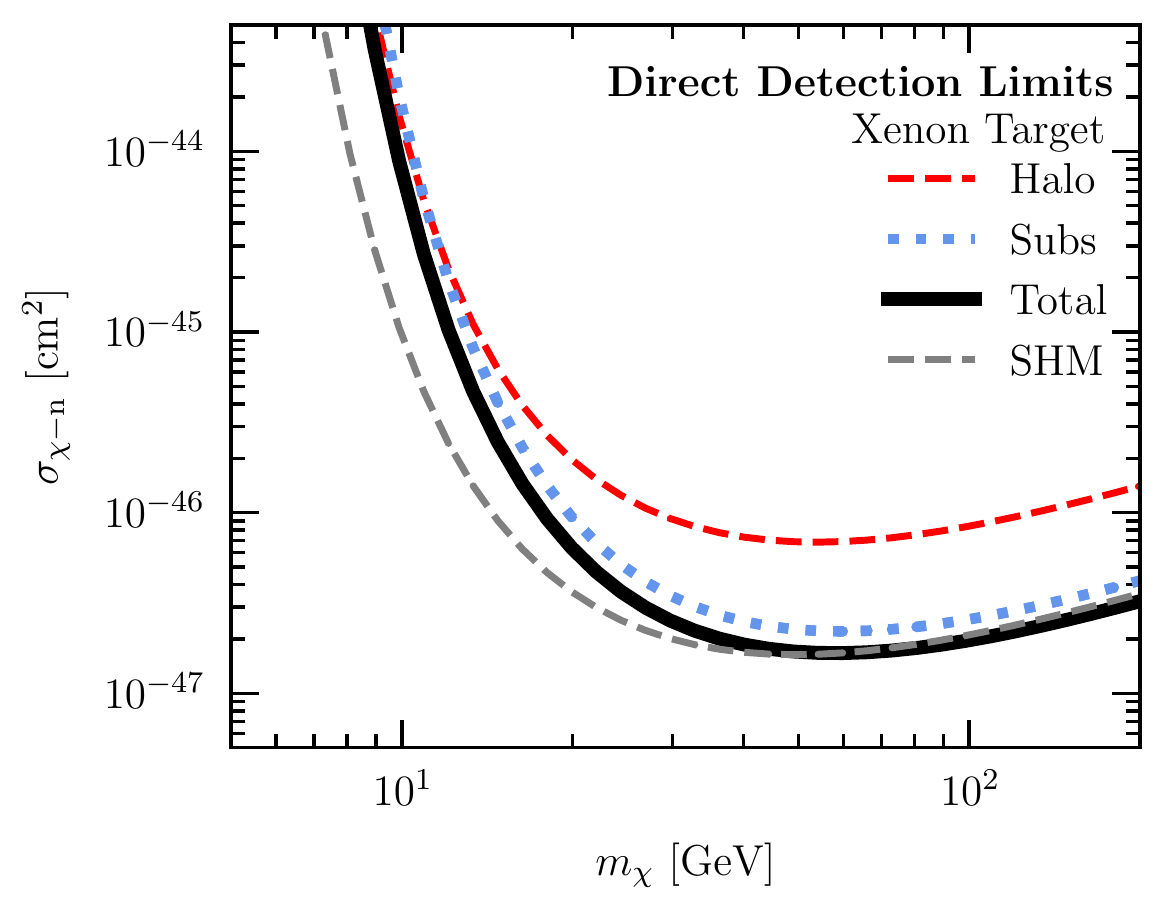}
\caption{(Left) Best-fit speed distribution for the halo (dashed red) and substructure (dotted blue) components. The solid black line represents the total contribution.  These results are based on fits to the SDSS-\emph{Gaia} DR2 data within heliocentric distances of $d_\odot < 4$~kpc and $|z| > 2.5$~kpc.  For comparison, we show the Standard Halo Model (dashed gray), defined in~\eqref{eq:shm}.  The empirical distribution does not include contributions  from DM accreted from non-luminous satellites or diffusely. 
(Right)~The 95\% background-free C.L. limits on the DM-nucleon scattering cross section, $\sigma_{\chi-n}$, for spin-independent interactions as a function of DM mass, $m_\chi$, assuming a xenon target with an exposure of 1 kton$\times$year exposure and a 4.9~keV$_\mathrm{nr}$ energy threshold. These limits are illustrative and do not account for experimental energy efficiencies near threshold~\citep{Aprile:2018dbl}. Note that the `Total' distribution assumes that the relative ratio of dark matter to stars contributed by accreted satellites in the halo and substructure populations is comparable. This assumption is further addressed in \cite{Necib:2018}. }
\label{fig:heliocentric}
\end{figure*}

In this section, we derive the heliocentric speed distribution for the halo and substructure populations, which we assume to trace the DM, and use it to calculate the scattering rate of a DM particle off a nuclear target.  Ideally, we need the stellar velocity distribution at the Solar position, but this is also the region where the disk contribution dominates. Any mis-modeling of the disk may therefore strongly bias the fit results in this regime.  For this reason, we restrict ourselves to $|z|>2.5$~kpc.  
The results from Sec.~3 give us confidence that the halo and substructure distributions are  likely invariant in $z$ and that we can extrapolate them into the plane, however this should be verified explicitly.  Previous work~\citep{2018MNRAS.478..611B} also finds evidence for the radial substructure down to $|z|= 1$~kpc.

The best-fit speed distribution in the heliocentric frame is shown in the left panel of Fig.~\ref{fig:heliocentric}, for heliocentric distance of $d_\odot < 4$~kpc and $|z| > 2.5$~kpc.
To obtain this distribution, we draw values of $v_r, v_\theta,$ and $v_\phi$ from the full posterior distribution in the Galactic frame.  Assuming that the stars are spatially uniform, we then transform to the heliocentric frame using the local rest-frame velocities $v_{\odot, \mathrm{pec}} =(U, V, W)$ = ($8.50$, $13.38$, $6.49$)~km/s~\citep{LSR} and local circular velocity $v_c = 235$~km/s. The substructure and halo components are plotted separately (blue dotted and red dashed, respectively).  
An important subtlety arises when summing these two contributions, as we need to know the relative amount of DM that is contributed by the accreted satellites in each population---\emph{i.e.,} the fractions $\xi_{1,2}$ in \Eq{eq: twocomponent}.  For now, we make the simplifying assumption that the satellites have comparable mass-to-light ratios.  In this case, the total contribution is shown as the solid black line.~\cite{Necib:2018} provide a more detailed prescription to estimate the relative dark matter fractions associated with each population.    

For comparison, we also plot the Standard Halo Model~(SHM) as the dashed gray line. The SHM is the speed distribution associated with the Maxwell Boltzmann (after integrating over the angular coordinates) and is defined as
\begin{equation}
f_\mathrm{SHM} (v) = \frac{4 v^2}{\sqrt{\pi} v_c^3} \exp \left[ - \frac{v^2}{v_c^2} \right] \, .
\label{eq:shm}
\end{equation}   
The dispersion of the SHM is closest to that of the halo posterior, albeit slightly higher; the SHM is isotropic with $\sigma \sim156~$km/s, while the best fits for the halo are $(\sigma_r, \sigma_\theta, \sigma_\phi)  =  \left( 140.3^{+4.2}_{-4.9}, 114.2^{+3.3}_{-1.8}, 125.9^{+4.1}_{-3.4}\right)$~km/s.  The primary discrepancy with the SHM arises from the substructure population.  When this component is included, the total speed distribution is discrepant with the SHM.  In this case, the polar and azimuthal velocities of the substructure are Gaussian with means $(\mu_\theta, \mu_\phi) = (-3.1^{+0.9}_{-0.9}, 35.5^{+1.8}_{-1.8})$~km/s and dispersions $(\sigma_\theta, \sigma_\phi)  = (57.7^{+0.7}_{-0.8}, 61.2^{+1.5}_{-1.5} )$~km/s, but the radial distribution has peaks at $\pm 117.7^{+1.8}_{-2.1}$~km/s, with a dispersion $\sigma_r = 108.2^{+1.2}_{-1.3}$~km/s for each. The substructure component arises from a more recent merger, as underlined by the fact that it is more metal-rich and highly radial.  Because it is not necessarily in steady state, there is no reason why the SHM should model it well.  

The empirical distribution clearly underestimates the fraction of high-speed DM particles, as compared to the SHM, which affects DM models where the minimum scattering speed, $v_\text{min}$, needed to create a nuclear recoil of energy $E_\text{nr}$ is high.  For elastic scattering, the minimum speed depends both on the DM particle properties as well as the experiment as follows:
\begin{equation}
v_\text{min} = \sqrt{\frac{m_N E_\mathrm{nr}}{2 \mu^2}} \, ,
\label{eq:vmin}
\end{equation}
where $m_\mathrm{N}$ is the nuclear mass and $\mu$ is the DM-nucleus reduced mass.  If the scattering is inelastic, then the minimum speed is even larger.  The differential scattering rate, per unit detector mass, for the most common operators is then
\begin{equation}
\frac{dR}{dE_\text{nr}} = \frac{\rho_\chi}{2 m_\chi \mu^2}  \, \sigma(q) \,  g\left(v_\mathrm{min}\right) \, ,
\label{eq:rate}
\end{equation}
where $m_\chi$ is the DM particle mass and $\rho_\chi$ its local density, $\sigma(q)$ is an effective scattering cross section that depends on the momentum transfer $q$, and the mean inverse speed is defined as 
\begin{equation}
g(v_\mathrm{min}) = \int_{v_\mathrm{min}}^{\infty} \frac{\tilde{f}(v)}{v} \, dv \, ,
\label{eq:gvmin}
\end{equation}
where $\tilde{f}(v)$ is the heliocentric velocity distribution.
\begin{figure*}[t] 
\centering
\includegraphics[width=0.45\textwidth]{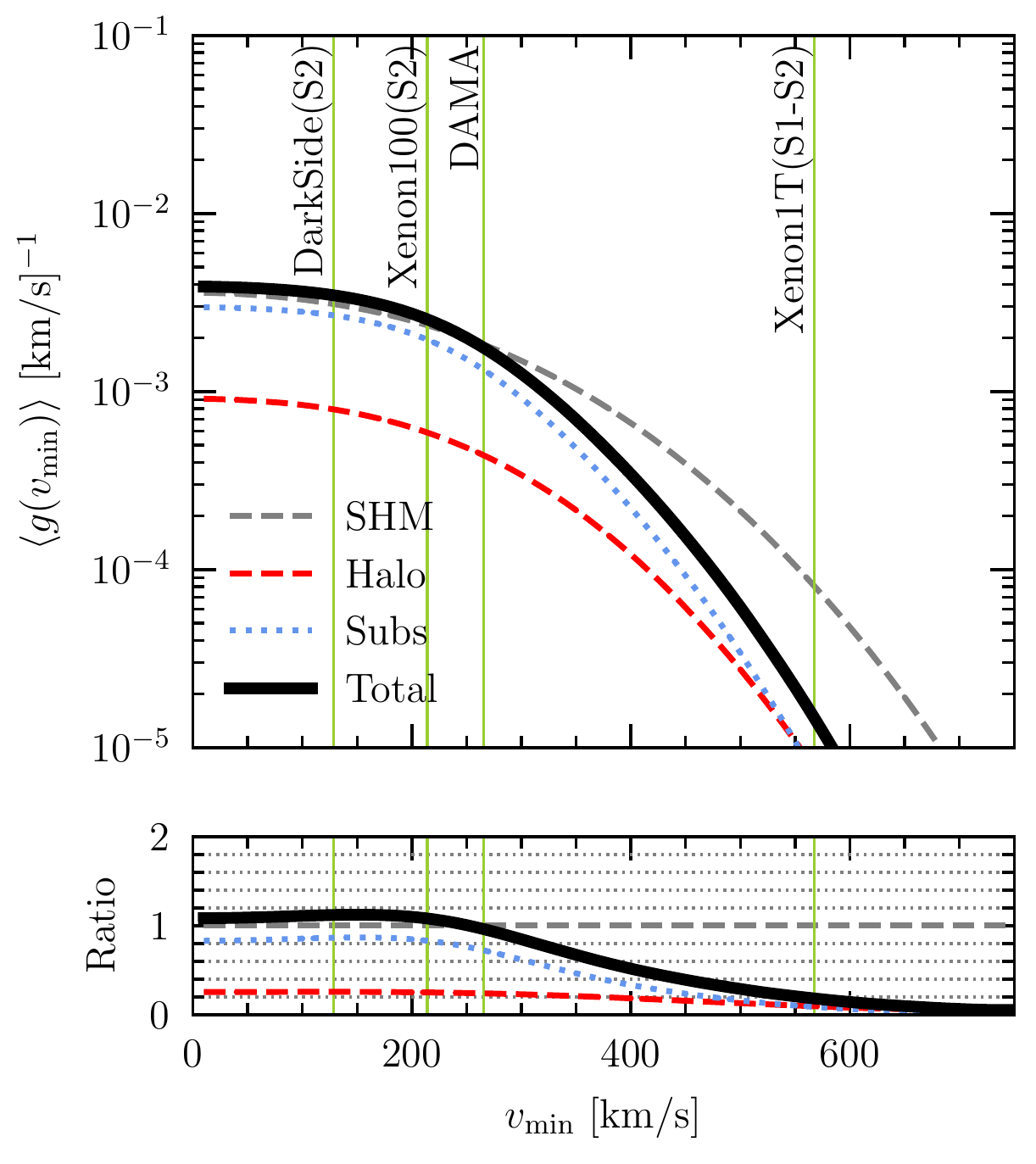}
\includegraphics[width=0.45\textwidth]{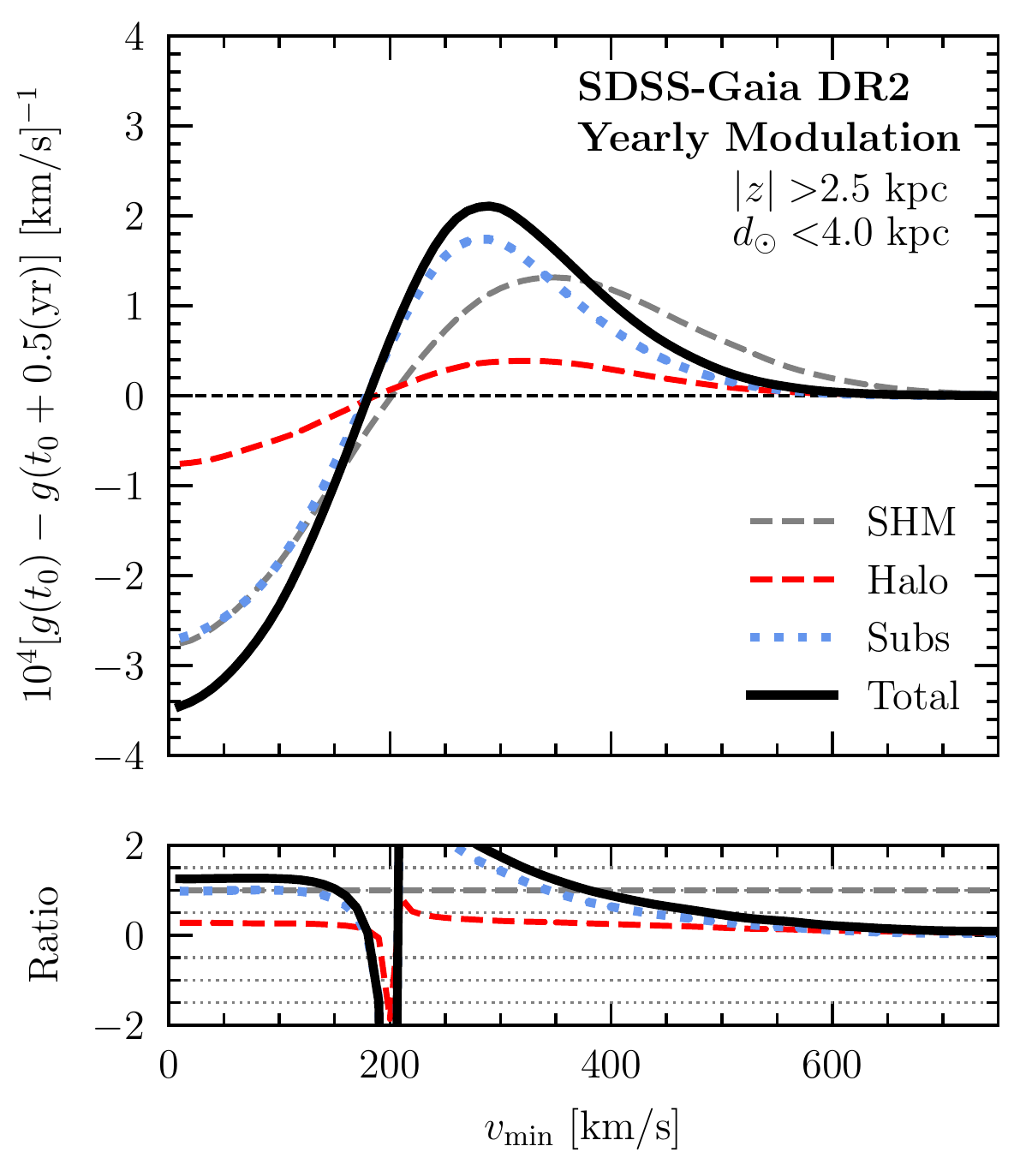}
\caption{(Left) Time-averaged inverse speed distribution defined in~\eqref{eq:gvmin} as a function of the minimum scattering speed, $v_\mathrm{min}$.  The substructure and halo distributions are shown as the dotted blue and dashed red lines, respectively, and their total contribution is shown as solid black.  The SHM expectation is the dashed gray line.  The vertical green lines indicate the values of $v_\mathrm{min}$ near threshold for a 10~GeV dark matter particle scattering in the DarkSide-50~\citep{Agnes:2018ves}, DAMA~\citep{Bernabei:2018yyw}, and Xenon1T~\citep{Aprile:2018dbl} detectors.  (Right) Expected yearly modulation amplitude between June ($t_0 \approx150$~days) and December as a function of $v_{\rm{min}}$.    Note that the `Total' distribution assumes that the relative ratio of dark matter to stars contributed by accreted satellites in the halo and substructure populations is comparable. This assumption is further addressed in \cite{Necib:2018}. }
\label{fig:gvmin}
\end{figure*}

The right panel of Fig.~\ref{fig:heliocentric} shows the corresponding limits on the DM mass and DM-nucleon scattering cross section, $\sigma_{\chi-n}$, assuming the simplest spin-independent operator.  For this example, we assume a xenon target, energy threshold of 4.9 keVnr, and exposure of 1~kton$\times$year. The 95\% one-sided Poisson C.L. limit (3~events) obtained using the velocity distribution inferred from SDSS-\emph{Gaia}~DR2 is shown in solid black, and compared to the SHM in dashed grey.  The substructure component drives the sensitivity at all masses, while the halo contribution is subdominant, but becomes more important at lower masses.  In both cases, the exclusion is  significantly weakened for $m_\chi \lesssim 30$~GeV relative to that obtained using the SHM.  For $m_\chi \gtrsim 100$~GeV, the black and gray-dashed lines approach each other because $v_\mathrm{min} \rightarrow 0$ in~\eqref{eq:gvmin}.

\begin{figure*}[t] 
\centering
\includegraphics[width=0.45\textwidth]{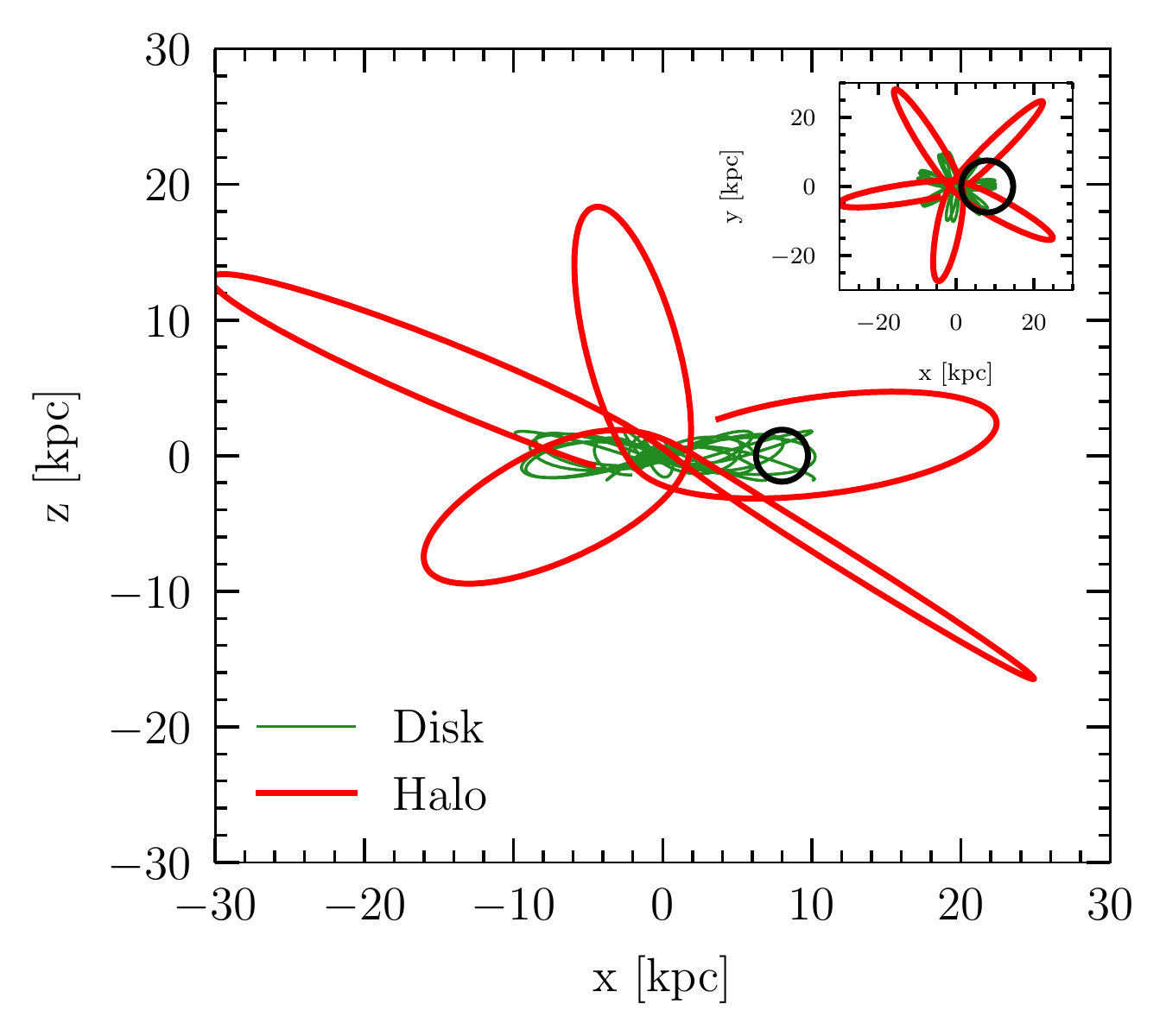}
\quad
\includegraphics[width=0.45\textwidth]{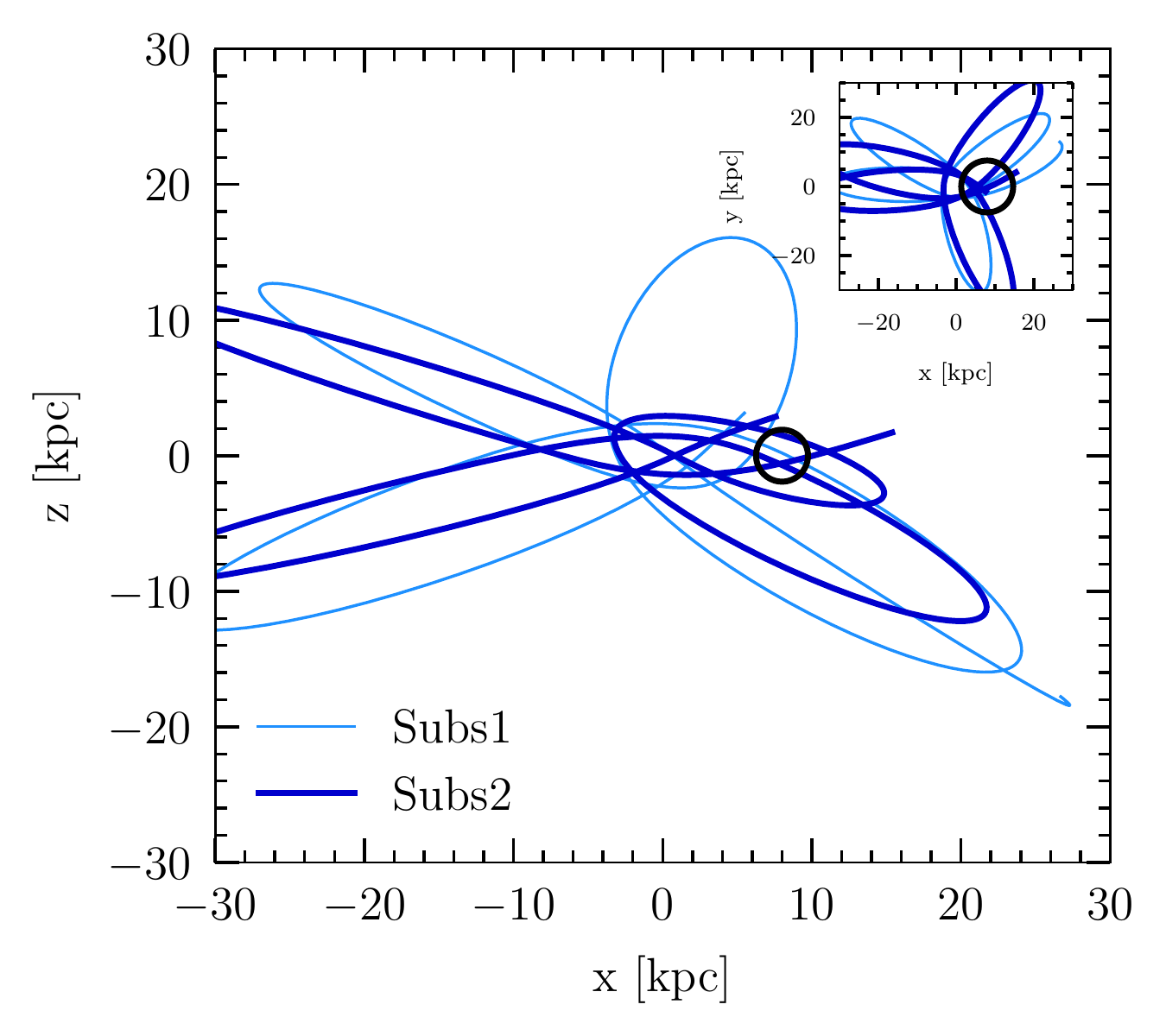}
\caption{(Left) Orbits of stars that likely belong to the disk (green) and halo (red) populations.  (Right) The corresponding orbits for two likely substructure stars, labeled as  `Subs1' and `Subs2.' The main figure shows the projection in the $x-z$ plane, while the inset shows that of the $x-y$ plane. The black circle shows the location of the Sun.}
\label{fig:orbits}
\end{figure*}

The overall effect of the empirical velocity distribution on the scattering limit depends on the details of the nuclear target, experimental threshold, and DM mass---all parameters that feed into the minimum scattering speed defined in~\eqref{eq:vmin}.  A more model- and experiment-independent way of understanding these effects is to study the dependence of the time-averaged inverse-speed, $\langle g(v_\mathrm{min}) \rangle$, as a function of the minimum speed, as this term captures the dependence of the scattering rate on the DM velocities.  The left panel of Fig.~\ref{fig:gvmin} plots this quantity for the empirical speed distribution obtained in this work (solid black) and the SHM (dashed gray).  The scattering rate for the empirical distribution is reduced relative to that for the SHM at $v_\text{min} \gtrsim 300$~km/s; it is enhanced for lower minimum speeds.  The scattering rate is completely suppressed for $v_\text{min} \gtrsim 550$~km/s, whereas the SHM continues to contribute events above this point.

To better understand the implications of these results, let us consider the concrete example of a 10~GeV DM particle interacting in several detectors.  Such a DM particle needs a minimum speed of $\sim 570$~km/s to scatter a xenon nucleus at an energy of $\sim5$~keV$_\text{nr}$ in Xenon1T~\citep{Aprile:2018dbl}  when requiring both S1 and S2 signals.\footnote{The S1 signal is associated with scintillation light from the initial particle collision, while the S2 signal is the scintillation light from the ionization electrons.}   As seen from the left panel of Fig.~\ref{fig:gvmin}, this is highly suppressed relative to the SHM expectation.\footnote{In actuality, Xenon1T has non-zero efficiency below $\sim~5$~keV$_\text{nr}$, which improves its sensitivity in this range.  }  In contrast, Xenon100 can detect recoils down to 0.7~keV$_{\rm{nr}}$ in the S2-only analysis \citep{Aprile:2016wwo}, similarly to the DarkSide-50 low-mass analysis~\citep{Agnes:2018ves}, which can detect argon recoils down to 0.6~keV$_\text{nr}$ in energy.  A 10~GeV DM particle only needs speeds of $\sim 130$--200~km/s to create such a recoil and these speeds are well-supported by the empirical distribution.

The empirical velocity distribution also impacts the time-dependence of a signal.  
The DM scattering rate should modulate annually 
due to the Earth's motion around the Sun~\citep{Drukier:1986tm}.  The right panel of \Fig{fig:gvmin} compares the modulation amplitude assuming the newly derived velocity distribution, as compared to the SHM.  To obtain the amplitude, we transform the velocities from the Galactic to the heliocentric frame, taking into account the Earth's time-dependent velocity as defined in \cite{Lee:2013xxa}.  We do not include the effect of gravitational focusing, which may further affect the properties of the modulation signal~\citep{Lee:2013wza}.

The modulation amplitude for the SHM exhibits the expected features: a maximum when $v_\mathrm{min} \sim 350$~km/s and a change in phase below $\sim 200$~km/s---see~\cite{Freese:2012xd} for a review.  In comparison, the  modulation amplitude obtained from the SDSS-\emph{Gaia}~DR2 distribution is maximal closer to $v_\mathrm{min} \sim 250$~km/s   and falls off faster towards higher $v_\mathrm{min}$.  This is due to the fact that the empirical velocity distribution $f(v)$ is less broad than the SHM.  Therefore, the differences in the heliocentric speed distribution over the year are typically more pronounced, but over a smaller range of $v_\text{min}$.   
The DAMA experiment, which claims an annually modulating signal, has a NaI target and threshold energy of $\sim3.3$~keV$_\text{nr}$~\citep{Bernabei:2018yyw}.  An observable scatter of a 10~GeV DM particle off a Na nucleus requires a minimum speed of $\sim270$~km/s.  We note that this falls in the region where the empirical distribution has a significant effect on the modulation amplitude, and motivates a more careful study of the consistency with data.

The results shown here are specific to spin-independent interactions.  However, the new velocity distribution will have an effect on other interaction operators---as the dependence of some of these operators on the DM momentum is non-trivial, the magnitude of the effects can vary from operator to operator~\citep{Fan:2010gt,Fitzpatrick:2012ix,Fitzpatrick:2012ib, Lisanti:2016jxe}.  In general, any model that relies on the high-velocity tail of the velocity distribution to support the scattering rate will be affected by these results.  One simple and illustrative case is that of inelastic (rather than elastic) spin-independent scattering off of nuclei.  Let us return to the example of the 10 GeV DM particle scattering off a xenon target at threshold $\sim0.7$~keV$_\text{nr}$.  The minimum scattering velocity for elastic scattering is $\sim210$~km/s, which is well supported by the velocity distribution as demonstrated in \Fig{fig:gvmin}.  However, if one considers a case where there are nearly degenerate states separated by $\sim15$~keV in mass, then the inelastic scattering requires $v_\text{min}\sim560$~km/s, which is suppressed relative to the SHM expectation.

Similarly, the velocity distribution will also be relevant for the interpretation of DM-electron scattering interactions---see~\cite{Battaglieri:2017aum} for a review---and axion experiments~\citep{Ling:2004aj, Hoskins:2016svf, Sloan:2016aub, 2017NuPhB.915...10V, Millar:2017eoc, Foster:2017hbq}.

\section{Conclusions}
\label{sec:conclusions}

We performed a mixture model analysis on MS stars in the SDSS-\emph{Gaia}~DR2 catalog within the range $r \in [7.5, 10.0]$~kpc and $|z| > 2.5$	~kpc.  The full chemo-dynamic properties of the stars ($\mathbf{v}$, $\FeH$) were used to identify the populations most likely belonging to the disk and halo, as well as any potential kinematic substructure.  The velocities of the disk and halo stars were modeled as multivariate normal distributions, while the substructure component was given the freedom to scan non-Gaussian possibilities.  

The recovered disk, halo, and substructure populations have median metallicities of $\FeH = -0.8, -1.8$, and $-1.4$ respectively.  
The disk component acts as expected within the region studied here.  The halo component is very metal-poor and its velocity ellipsoid is nearly isotropic.  The analysis identifies a substructure population of intermediate metallicity stars whose radial velocities are best modeled with a non-Gaussian distribution.  This population had been identified in previous work as the \emph{Gaia} Sausage~\citep{2018MNRAS.478..611B}.  Our analysis provides the first model for its velocity and metallicity distribution, and clearly distinguishes its contribution relative to the disk and metal-poor halo over the full metallicity range of the sample.

The substructure population is anisotropic in velocity, with two broad lobes centered at $v_r \sim \pm 150$~km/s. 
It can be explained as tidal debris from a satellite galaxy on a highly radial orbit.  The lobes are consistent with debris that is torn off as the satellite moves towards/away from the Galactic Center while orbiting.  The distinctive metallicity of the substructure strongly suggests that it is sourced by a single progenitor.  

To illustrate these points, Fig.~\ref{fig:orbits} compares the orbits of a likely halo, disk, and substructure star, chosen at random---see also~\cite{2018arXiv180510288D}.  We use the \texttt{gala} package~\citep{PriceWhelan:2017} to integrate the orbits back 1~Myr, given the star's present-day position and velocity, and assume the default Milky Way potential from~\cite{2015ApJS..216...29B}.  In the left panel, we plot the orbit of a likely disk (green line) and halo (red line) star.  The disk star is confined to the plane, as expected, while the halo star's orbit is more isotropic.  We contrast this to the orbits of two likely substructure stars in the right panel, indicated by the dark and light blue lines.  The orbits of these stars are highly radial and have a lower inclination angle relative to the mid-plane.  

As the substructure is fairly hot and exhibits no obvious spatial features in the local region studied here, it likely originated from an old merger.  Using the stellar-mass metallicity relation of~\cite{Kirby:2013wna} and the median metallicity of the substructure component, we estimate that its progenitor had stellar mass $M_* \sim 10^{7-8}$~M$_\odot$.  We note that the Sagittarius dwarf galaxy~\citep{Ibata:1994fv} cannot be the progenitor of the substructure as it is a younger merger event and its orbit is less eccentric~\citep{Law:2010pe, Purcell:2011nf}.  In particular, the apocenter to pericenter ratio of Sagittarius is 5:1, whereas it is 20:1 for the substructure.

Our study of the local stellar distribution has direct relevance for DM.  Numerical simulations have demonstrated that the old metal-poor halo is a good tracer of the virialized DM kinematics.  Additionally, kinematic substructure such as DM debris flow have been shown to have stellar counterparts.  If the accreted stellar component in the SDSS-\emph{Gaia}~DR2 sample (\emph{e.g.}, the halo and substructure populations) are adequate DM tracers, then our results imply that a fraction of the local DM is in debris flow. 

We find that the halo and substructure populations do not depend on the vertical distance off the plane, at least in the region from $|z| > 2.5$--$4$~kpc.  This gives us confidence in extrapolating their contributions to the Solar neighborhood, which is relevant for direct detection experiments.  By performing a more detailed modeling of the disk component, one could potentially extend the mixture analysis to lower $z$ and recover the accreted stellar distribution directly in this region.  We plan to pursue this in future work.  

The heliocentric speed distribution that we derive from SDSS-\emph{Gaia}~DR2 within heliocentric distances $d_\odot < 4$~kpc and $|z| > 2.5$~kpc is incompatible with the SHM.  The inferred DM distribution has far fewer high-speed particles than expected from the SHM.  If we assume that the distributions recovered in this study extrapolate down to the Galactic mid-plane, then this would reduce the sensitivity to DM that is not energetic enough to create observable nuclear recoils in a detector target.  In this case, the limits of an experiment using a xenon target are suppressed below $m_\mathrm{DM}\lesssim 30$~GeV for the case of spin-independent interactions.  The overall size of the suppression can vary for different nuclear target masses, as well as different scattering operators.  Current exclusion limits and future projections should be revisited in light of these new findings.  

The SDSS-\emph{Gaia} DR2 study provides the first indication that a subset of the local DM may be in kinematic substructure.
Given these results, it is pressing to better quantify just how well the stars and DM track each other in simulated mergers that resemble the observations.  Additionally, we need to quantify the effects of DM that is diffuse or originates from non-luminous subhalos.  If either of these dominates locally, then the total DM distribution will differ from that accreted by the largest satellites.  Furthermore, improved modeling is needed to better estimate the relative dark matter fraction contributed by the accreted satellites in the halo and substructure populations.   In particular, \cite{Necib:2018} presents an empirical approach to estimate the relative DM contribution from each satellite, by inferring the average metallicity of the accreted stars of a merger and the ratio of DM mass to stellar mass of the merging satellite.  Using such a relationship, they estimated that the $42^{+26}_{-22}\%$ of DM accreted from luminous satellites is in debris flow.

\section*{Acknowledgements}

We would like to thank J.~Bochanski, J.~Bovy, G.~Collin, A.~Drlica-Wagner, D.~Hogg, K.~Hawkins, A.~Kaboth, L.~Lancaster, T.~Li, S.~McDermott, G~C.~Myeong, D.~Spergel, and N.~Weiner for useful conversations.  We also acknowledge use of the \texttt{Astropy}~\citep{2013A&A...558A..33A}, \texttt{emcee}~\citep{2013PASP..125..306F}, \texttt{gala}~\citep{PriceWhelan:2017}, and \texttt{IPython}~\citep{PER-GRA:2007} software packages.  This project was developed in part at the NYC Gaia DR2 Workshop in 2018 April, as well as the 2018 NYC Gaia Sprint, hosted by the Center for Computational Astrophysics of the Flatiron Institute in New York City.

LN is supported by the DOE under Award Number DESC0011632, and the Sherman Fairchild fellowship. ML is supported by the DOE under Award Number DESC0007968, the Alfred P. Sloan Foundation and the Cottrell Scholar Program through the Research Corporation for Science Advancement.  VB is supported by the European Research Council under the European Union's Seventh
Framework Programme (FP/2007-2013) / ERC Grant Agreement
n. 308024. This paper made used of the Whole Sky Database (WSDB)
created by Sergey Koposov and maintained at the Institute of
Astronomy, Cambridge by Sergey Koposov, Vasily Belokurov and Wyn Evans
with financial support from the Science \& Technology Facilities
Council (STFC) and the European Research Council (ERC).

This work has made use of data from the European Space Agency (ESA)
mission {\it Gaia} (\url{https://www.cosmos.esa.int/gaia}), processed by
the {\it Gaia} Data Processing and Analysis Consortium (DPAC,
\url{https://www.cosmos.esa.int/web/gaia/dpac/consortium}). Funding
for the DPAC has been provided by national institutions, in particular
the institutions participating in the {\it Gaia} Multilateral Agreement.  

Funding for SDSS-III has been provided by the Alfred P. Sloan Foundation, the Participating Institutions, the National Science Foundation, and the U.S. Department of Energy Office of Science. The SDSS-III web site is \url{http://www.sdss3.org/}.

SDSS-III is managed by the Astrophysical Research Consortium for the Participating Institutions of the SDSS-III Collaboration including the University of Arizona, the Brazilian Participation Group, Brookhaven National Laboratory, Carnegie Mellon University, University of Florida, the French Participation Group, the German Participation Group, Harvard University, the Instituto de Astrofisica de Canarias, the Michigan State/Notre Dame/JINA Participation Group, Johns Hopkins University, Lawrence Berkeley National Laboratory, Max Planck Institute for Astrophysics, Max Planck Institute for Extraterrestrial Physics, New Mexico State University, New York University, Ohio State University, Pennsylvania State University, University of Portsmouth, Princeton University, the Spanish Participation Group, University of Tokyo, University of Utah, Vanderbilt University, University of Virginia, University of Washington, and Yale University.

\clearpage
\def\bibsection{} 
\bibliographystyle{aasjournal}
\bibliography{sdss}

\begin{thebibliography}{}
\expandafter\ifx\csname natexlab\endcsname\relax\def\natexlab#1{#1}\fi

\bibitem[{Agnes {et~al.}(2018)}]{Agnes:2018ves}
Agnes, P., {et~al.} 2018, arXiv:1802.06994

\bibitem[{{Ahn} {et~al.}(2012){Ahn}, {Alexandroff}, {Allende Prieto},
  {Anderson}, {Anderton}, {Andrews}, {Aubourg}, {Bailey}, {Balbinot}, {Barnes},
  \& et~al.}]{2012ApJS..203...21A}
{Ahn}, C.~P., {Alexandroff}, R., {Allende Prieto}, C., {et~al.} 2012, \apjs,
  203, 21

\bibitem[{Aprile {et~al.}(2016)}]{Aprile:2016wwo}
Aprile, E., {et~al.} 2016, Phys. Rev., D94, 092001, [Erratum: Phys.
  Rev.D95,no.5,059901(2017)]

\bibitem[{Aprile {et~al.}(2018)}]{Aprile:2018dbl}
---. 2018, arXiv:1805.12562

\bibitem[{{Astropy Collaboration} {et~al.}(2013){Astropy Collaboration},
  {Robitaille}, {Tollerud}, {Greenfield}, {Droettboom}, {Bray}, {Aldcroft},
  {Davis}, {Ginsburg}, {Price-Whelan}, {Kerzendorf}, {Conley}, {Crighton},
  {Barbary}, {Muna}, {Ferguson}, {Grollier}, {Parikh}, {Nair}, {Unther},
  {Deil}, {Woillez}, {Conseil}, {Kramer}, {Turner}, {Singer}, {Fox}, {Weaver},
  {Zabalza}, {Edwards}, {Azalee Bostroem}, {Burke}, {Casey}, {Crawford},
  {Dencheva}, {Ely}, {Jenness}, {Labrie}, {Lim}, {Pierfederici}, {Pontzen},
  {Ptak}, {Refsdal}, {Servillat}, \& {Streicher}}]{2013A&A...558A..33A}
{Astropy Collaboration}, {Robitaille}, T.~P., {Tollerud}, E.~J., {et~al.} 2013,
  AAP, 558, A33

\bibitem[{{Bahcall} \& {Soneira}(1980)}]{1980ApJS...44...73B}
{Bahcall}, J.~N., \& {Soneira}, R.~M. 1980, \apjs, 44, 73

\bibitem[{Battaglieri {et~al.}(2017)}]{Battaglieri:2017aum}
Battaglieri, M., {et~al.} 2017, arXiv:1707.04591

\bibitem[{{Belokurov} {et~al.}(2018{\natexlab{a}}){Belokurov}, {Deason},
  {Koposov}, {Catelan}, {Erkal}, {Drake}, \& {Evans}}]{2018MNRAS.477.1472B}
{Belokurov}, V., {Deason}, A.~J., {Koposov}, S.~E., {et~al.}
  2018{\natexlab{a}}, \mnras, 477, 1472

\bibitem[{{Belokurov} {et~al.}(2018{\natexlab{b}}){Belokurov}, {Erkal},
  {Evans}, {Koposov}, \& {Deason}}]{2018MNRAS.478..611B}
{Belokurov}, V., {Erkal}, D., {Evans}, N.~W., {Koposov}, S.~E., \& {Deason},
  A.~J. 2018{\natexlab{b}}, \mnras, 478, 611

\bibitem[{Bernabei {et~al.}(2018)}]{Bernabei:2018yyw}
Bernabei, R., {et~al.} 2018, arXiv:1805.10486

\bibitem[{Bhattacharjee {et~al.}(2013)Bhattacharjee, Chaudhury, Kundu, \&
  Majumdar}]{Bhattacharjee:2012xm}
Bhattacharjee, P., Chaudhury, S., Kundu, S., \& Majumdar, S. 2013, Phys. Rev.,
  D87, 083525

\bibitem[{{Blitz}(1979)}]{1979ApJ...231L.115B}
{Blitz}, L. 1979, \apjl, 231, L115

\bibitem[{Bond {et~al.}(2010)}]{Bond:2009mh}
Bond, N.~A., {et~al.} 2010, Astrophys. J., 716, 1

\bibitem[{{Bovy}(2015)}]{2015ApJS..216...29B}
{Bovy}, J. 2015, The Astrophysical Journal Supplement Series, 216, 29

\bibitem[{Bozorgnia \& Bertone(2017)}]{Bozorgnia:2017brl}
Bozorgnia, N., \& Bertone, G. 2017, arXiv:1705.05853

\bibitem[{{Bozorgnia} {et~al.}(2013){Bozorgnia}, {Catena}, \&
  {Schwetz}}]{2013JCAP...12..050B}
{Bozorgnia}, N., {Catena}, R., \& {Schwetz}, T. 2013, \jcap, 12, 050

\bibitem[{Bozorgnia {et~al.}(2016)}]{Bozorgnia:2016ogo}
Bozorgnia, N., {et~al.} 2016, JCAP, 1605, 024

\bibitem[{{Burton} \& {Gordon}(1978)}]{1978A&A....63....7B}
{Burton}, W.~B., \& {Gordon}, M.~A. 1978, \aap, 63, 7

\bibitem[{Caldwell \& Ostriker(1981)}]{Caldwell:1981rj}
Caldwell, J. A.~R., \& Ostriker, J.~P. 1981, Astrophys. J., 251, 61

\bibitem[{Carollo {et~al.}(2007)}]{Carollo:2007xh}
Carollo, D., {et~al.} 2007, Nature, 450, 1020

\bibitem[{Carollo {et~al.}(2010)Carollo, Beers, Chiba, Norris, Freeman, Lee,
  Ivezic, Rockosi, \& Yanny}]{Carollo:2009dz}
Carollo, D., Beers, T.~C., Chiba, M., {et~al.} 2010, Astrophys. J., 712, 692

\bibitem[{{Catena} \& {Ullio}(2012)}]{2012JCAP...05..005C}
{Catena}, R., \& {Ullio}, P. 2012, \jcap, 5, 005

\bibitem[{Chaudhury {et~al.}(2010)Chaudhury, Bhattacharjee, \&
  Cowsik}]{Chaudhury:2010hj}
Chaudhury, S., Bhattacharjee, P., \& Cowsik, R. 2010, JCAP, 1009, 020

\bibitem[{{Clemens}(1985)}]{1985ApJ...295..422C}
{Clemens}, D.~P. 1985, \apj, 295, 422

\bibitem[{{Co{\c s}kuno{\v g}lu} {et~al.}(2011){Co{\c s}kuno{\v g}lu}, {Ak},
  {Bilir}, {Karaali}, {Yaz}, {Gilmore}, {Seabroke}, {Bienaym{\'e}},
  {Bland-Hawthorn}, {Campbell}, {Freeman}, {Gibson}, {Grebel}, {Munari},
  {Navarro}, {Parker}, {Siebert}, {Siviero}, {Steinmetz}, {Watson}, {Wyse}, \&
  {Zwitter}}]{LSR}
{Co{\c s}kuno{\v g}lu}, B., {Ak}, S., {Bilir}, S., {et~al.} 2011, \mnras, 412,
  1237

\bibitem[{Deason {et~al.}(2013)Deason, Belokurov, Evans, \&
  Johnston}]{Deason:2012fc}
Deason, A.~J., Belokurov, V., Evans, N.~W., \& Johnston, K.~V. 2013, Astrophys.
  J., 763, 113

\bibitem[{{Deason} {et~al.}(2018){Deason}, {Belokurov}, {Koposov}, \&
  {Lancaster}}]{2018arXiv180510288D}
{Deason}, A.~J., {Belokurov}, V., {Koposov}, S.~E., \& {Lancaster}, L. 2018,
  ArXiv e-prints, arXiv:1805.10288

\bibitem[{{Deason} {et~al.}(2015){Deason}, {Belokurov}, \&
  {Weisz}}]{2015MNRAS.448L..77D}
{Deason}, A.~J., {Belokurov}, V., \& {Weisz}, D.~R. 2015, \mnras, 448, L77

\bibitem[{Del~Nobile(2014)}]{Cremonesi:2013bma}
Del~Nobile, E. 2014, Adv. High Energy Phys., 2014, 604914

\bibitem[{Diemand {et~al.}(2007)Diemand, Kuhlen, \& Madau}]{Diemand:2007qr}
Diemand, J., Kuhlen, M., \& Madau, P. 2007, Astrophys. J., 667, 859

\bibitem[{{Diemand} {et~al.}(2008){Diemand}, {Kuhlen}, {Madau}, {Zemp},
  {Moore}, {Potter}, \& {Stadel}}]{2008Natur.454..735D}
{Diemand}, J., {Kuhlen}, M., {Madau}, P., {et~al.} 2008, \nat, 454, 735

\bibitem[{Drukier {et~al.}(1986)Drukier, Freese, \& Spergel}]{Drukier:1986tm}
Drukier, A.~K., Freese, K., \& Spergel, D.~N. 1986, Phys. Rev., D33, 3495

\bibitem[{Elahi {et~al.}(2011)Elahi, Thacker, \& Widrow}]{Elahi:2011dy}
Elahi, P.~J., Thacker, R.~J., \& Widrow, L.~M. 2011, Mon.Not.Roy.Astron.Soc.,
  418, 320

\bibitem[{{ESA}(1997)}]{1997ESASP1200.....E}
{ESA}, ed. 1997, ESA Special Publication, Vol. 1200, {The HIPPARCOS and TYCHO
  catalogues. Astrometric and photometric star catalogues derived from the ESA
  HIPPARCOS Space Astrometry Mission}

\bibitem[{Fairbairn \& Schwetz(2009)}]{Fairbairn:2008gz}
Fairbairn, M., \& Schwetz, T. 2009, JCAP, 0901, 037

\bibitem[{Fan {et~al.}(2010)Fan, Reece, \& Wang}]{Fan:2010gt}
Fan, J., Reece, M., \& Wang, L.-T. 2010, JCAP, 1011, 042

\bibitem[{{Fich} {et~al.}(1989){Fich}, {Blitz}, \&
  {Stark}}]{1989ApJ...342..272F}
{Fich}, M., {Blitz}, L., \& {Stark}, A.~A. 1989, \apj, 342, 272

\bibitem[{{Fiorentino} {et~al.}(2015){Fiorentino}, {Bono}, {Monelli},
  {Stetson}, {Tolstoy}, {Gallart}, {Salaris}, {Mart{\'\i}nez-V{\'a}zquez}, \&
  {Bernard}}]{2015ApJ...798L..12F}
{Fiorentino}, G., {Bono}, G., {Monelli}, M., {et~al.} 2015, \apj, 798, L12

\bibitem[{Fitzpatrick {et~al.}(2012)Fitzpatrick, Haxton, Katz, Lubbers, \&
  Xu}]{Fitzpatrick:2012ib}
Fitzpatrick, A.~L., Haxton, W., Katz, E., Lubbers, N., \& Xu, Y. 2012,
  arXiv:1211.2818

\bibitem[{Fitzpatrick {et~al.}(2013)Fitzpatrick, Haxton, Katz, Lubbers, \&
  Xu}]{Fitzpatrick:2012ix}
---. 2013, JCAP, 1302, 004

\bibitem[{{Foreman-Mackey} {et~al.}(2013){Foreman-Mackey}, {Hogg}, {Lang}, \&
  {Goodman}}]{2013PASP..125..306F}
{Foreman-Mackey}, D., {Hogg}, D.~W., {Lang}, D., \& {Goodman}, J. 2013, \pasp,
  125, 306

\bibitem[{Fornasa \& Green(2014)}]{Fornasa:2013iaa}
Fornasa, M., \& Green, A.~M. 2014, Phys. Rev., D89, 063531

\bibitem[{Foster {et~al.}(2017)Foster, Rodd, \& Safdi}]{Foster:2017hbq}
Foster, J.~W., Rodd, N.~L., \& Safdi, B.~R. 2017, arXiv:1711.10489

\bibitem[{Freese {et~al.}(1988)Freese, Frieman, \& Gould}]{Freese:1987wu}
Freese, K., Frieman, J.~A., \& Gould, A. 1988, Phys. Rev., D37, 3388

\bibitem[{Freese {et~al.}(2013)Freese, Lisanti, \& Savage}]{Freese:2012xd}
Freese, K., Lisanti, M., \& Savage, C. 2013, Rev. Mod. Phys., 85, 1561

\bibitem[{{Gaia Collaboration} {et~al.}(2018){Gaia Collaboration}, {Brown},
  {Vallenari}, {Prusti}, {de Bruijne}, {Babusiaux}, \&
  {Bailer-Jones}}]{2018arXiv180409365G}
{Gaia Collaboration}, {Brown}, A.~G.~A., {Vallenari}, A., {et~al.} 2018, ArXiv
  e-prints, arXiv:1804.09365

\bibitem[{{Gaia Collaboration} {et~al.}(2016){Gaia Collaboration}, {Prusti},
  {de Bruijne}, {Brown}, {Vallenari}, {Babusiaux}, {Bailer-Jones}, {Bastian},
  {Biermann}, {Evans}, \& et~al.}]{2016A&A...595A...1G}
{Gaia Collaboration}, {Prusti}, T., {de Bruijne}, J.~H.~J., {et~al.} 2016,
  \aap, 595, A1

\bibitem[{Green(2017)}]{Green:2017odb}
Green, A.~M. 2017, J. Phys., G44, 084001

\bibitem[{Hansen \& Moore(2006)}]{Hansen:2004qs}
Hansen, S.~H., \& Moore, B. 2006, New Astron., 11, 333

\bibitem[{{Helmi}(2008)}]{2008A&ARv..15..145H}
{Helmi}, A. 2008, \aapr, 15, 145

\bibitem[{{Helmi} {et~al.}(2018){Helmi}, {Babusiaux}, {Koppelman}, {Massari},
  {Veljanoski}, \& {Brown}}]{2018arXiv180606038H}
{Helmi}, A., {Babusiaux}, C., {Koppelman}, H.~H., {et~al.} 2018, ArXiv
  e-prints, arXiv:1806.06038

\bibitem[{Helmi \& White(1999)}]{Helmi:1999ks}
Helmi, A., \& White, S. D.~M. 1999, Mon. Not. Roy. Astron. Soc., 307, 495

\bibitem[{Herzog-Arbeitman {et~al.}(2017{\natexlab{a}})Herzog-Arbeitman,
  Lisanti, Madau, \& Necib}]{eris_paper}
Herzog-Arbeitman, J., Lisanti, M., Madau, P., \& Necib, L. 2017{\natexlab{a}},
  arXiv:1704.04499

\bibitem[{Herzog-Arbeitman {et~al.}(2017{\natexlab{b}})Herzog-Arbeitman,
  Lisanti, \& Necib}]{Herzog-Arbeitman:2017zbm}
Herzog-Arbeitman, J., Lisanti, M., \& Necib, L. 2017{\natexlab{b}},
  arXiv:1708.03635

\bibitem[{Hoskins {et~al.}(2016)}]{Hoskins:2016svf}
Hoskins, J., {et~al.} 2016, Phys. Rev., D94, 082001

\bibitem[{Ibata {et~al.}(1994)Ibata, Gilmore, \& Irwin}]{Ibata:1994fv}
Ibata, R.~A., Gilmore, G., \& Irwin, M.~J. 1994, Nature, 370, 194

\bibitem[{Ivezic {et~al.}(2008)}]{Ivezic:2008wk}
Ivezic, Z., {et~al.} 2008, Astrophys. J., 684, 287

\bibitem[{{Jeans}(1922)}]{1922MNRAS..82..122J}
{Jeans}, J.~H. 1922, \mnras, 82, 122

\bibitem[{{Jungman} {et~al.}(1996){Jungman}, {Kamionkowski}, \&
  {Griest}}]{1996PhR...267..195J}
{Jungman}, G., {Kamionkowski}, M., \& {Griest}, K. 1996, \physrep, 267, 195

\bibitem[{{Kapteyn}(1922)}]{1922ApJ....55..302K}
{Kapteyn}, J.~C. 1922, \apj, 55, 302

\bibitem[{Kelso {et~al.}(2016)}]{Kelso:2016qqj}
Kelso, C., {et~al.} 2016, JCAP, 1608, 071

\bibitem[{Kirby {et~al.}(2013)Kirby, Cohen, Guhathakurta, Cheng, Bullock, \&
  Gallazzi}]{Kirby:2013wna}
Kirby, E.~N., Cohen, J.~G., Guhathakurta, P., {et~al.} 2013, Astrophys. J.,
  779, 102

\bibitem[{{Knapp} {et~al.}(1985){Knapp}, {Stark}, \&
  {Wilson}}]{1985AJ.....90..254K}
{Knapp}, G.~R., {Stark}, A.~A., \& {Wilson}, R.~W. 1985, \aj, 90, 254

\bibitem[{{Koposov} \& {Bartunov}(2006)}]{q3c}
{Koposov}, S., \& {Bartunov}, O. 2006, in Astronomical Society of the Pacific
  Conference Series, Vol. 351, Astronomical Data Analysis Software and Systems
  XV, ed. C.~{Gabriel}, C.~{Arviset}, D.~{Ponz}, \& S.~{Enrique}, 735

\bibitem[{Kuhlen {et~al.}(2012)Kuhlen, Lisanti, \& Spergel}]{Kuhlen:2012fz}
Kuhlen, M., Lisanti, M., \& Spergel, D.~N. 2012, Phys. Rev., D86, 063505

\bibitem[{Kuhlen {et~al.}(2010)Kuhlen, Weiner, Diemand, Madau, Moore, Potter,
  Stadel, \& Zemp}]{Kuhlen:2009vh}
Kuhlen, M., Weiner, N., Diemand, J., {et~al.} 2010, JCAP, 1002, 030

\bibitem[{{Kunder} {et~al.}(2017){Kunder}, {Kordopatis}, {Steinmetz},
  {Zwitter}, {McMillan}, {Casagrande}, {Enke}, {Wojno}, {Valentini},
  {Chiappini}, {Matijevi{\v c}}, {Siviero}, {de Laverny}, {Recio-Blanco},
  {Bijaoui}, {Wyse}, {Binney}, {Grebel}, {Helmi}, {Jofre}, {Antoja}, {Gilmore},
  {Siebert}, {Famaey}, {Bienaym{\'e}}, {Gibson}, {Freeman}, {Navarro},
  {Munari}, {Seabroke}, {Anguiano}, {{\v Z}erjal}, {Minchev}, {Reid},
  {Bland-Hawthorn}, {Kos}, {Sharma}, {Watson}, {Parker}, {Scholz}, {Burton},
  {Cass}, {Hartley}, {Fiegert}, {Stupar}, {Ritter}, {Hawkins}, {Gerhard},
  {Chaplin}, {Davies}, {Elsworth}, {Lund}, {Miglio}, \&
  {Mosser}}]{2017AJ....153...75K}
{Kunder}, A., {Kordopatis}, G., {Steinmetz}, M., {et~al.} 2017, \aj, 153, 75

\bibitem[{{Lancaster} {et~al.}(2018){Lancaster}, {Koposov}, {Belokurov},
  {Evans}, \& {Deason}}]{LancasterBHBs}
{Lancaster}, L., {Koposov}, S., {Belokurov}, V., {Evans}, N.~W., \& {Deason},
  A. 2018

\bibitem[{Law \& Majewski(2010)}]{Law:2010pe}
Law, D.~R., \& Majewski, S.~R. 2010, Astrophys. J., 714, 229

\bibitem[{Lee {et~al.}(2014)Lee, Lisanti, Peter, \& Safdi}]{Lee:2013wza}
Lee, S.~K., Lisanti, M., Peter, A. H.~G., \& Safdi, B.~R. 2014, Phys. Rev.
  Lett., 112, 011301

\bibitem[{Lee {et~al.}(2013)Lee, Lisanti, \& Safdi}]{Lee:2013xxa}
Lee, S.~K., Lisanti, M., \& Safdi, B.~R. 2013, JCAP, 1311, 033

\bibitem[{Ling {et~al.}(2010)Ling, Nezri, Athanassoula, \&
  Teyssier}]{Ling:2009eh}
Ling, F.~S., Nezri, E., Athanassoula, E., \& Teyssier, R. 2010, JCAP, 1002, 012

\bibitem[{Ling {et~al.}(2004)Ling, Sikivie, \& Wick}]{Ling:2004aj}
Ling, F.-S., Sikivie, P., \& Wick, S. 2004, Phys. Rev., D70, 123503

\bibitem[{Lisanti(2017)}]{Lisanti:2016jxe}
Lisanti, M. 2017, in {Proceedings, Theoretical Advanced Study Institute in
  Elementary Particle Physics: New Frontiers in Fields and Strings (TASI 2015):
  Boulder, CO, USA, June 1-26, 2015}, 399--446

\bibitem[{Lisanti \& Spergel(2012)}]{Lisanti:2011as}
Lisanti, M., \& Spergel, D.~N. 2012, Phys. Dark Univ., 1, 155

\bibitem[{Lisanti {et~al.}(2015)Lisanti, Spergel, \& Madau}]{Lisanti:2014dva}
Lisanti, M., Spergel, D.~N., \& Madau, P. 2015, Astrophys. J., 807, 14

\bibitem[{Lisanti {et~al.}(2011)Lisanti, Strigari, Wacker, \&
  Wechsler}]{Lisanti:2010qx}
Lisanti, M., Strigari, L.~E., Wacker, J.~G., \& Wechsler, R.~H. 2011, Phys.
  Rev., D83, 023519

\bibitem[{Maciejewski {et~al.}(2011)Maciejewski, Vogelsberger, White, \&
  Springel}]{Maciejewski:2010gz}
Maciejewski, M., Vogelsberger, M., White, S. D.~M., \& Springel, V. 2011, Mon.
  Not. Roy. Astron. Soc., 415, 2475

\bibitem[{Mandal {et~al.}(2018)Mandal, Majumdar, Rentala, \&
  Basu~Thakur}]{Mandal:2018efq}
Mandal, S., Majumdar, S., Rentala, V., \& Basu~Thakur, R. 2018,
  arXiv:1806.06872

\bibitem[{Mao {et~al.}(2013)Mao, Strigari, Wechsler, Wu, \& Hahn}]{Mao:2012hf}
Mao, Y.-Y., Strigari, L.~E., Wechsler, R.~H., Wu, H.-Y., \& Hahn, O. 2013,
  Astrophys. J., 764, 35

\bibitem[{March-Russell {et~al.}(2009)March-Russell, McCabe, \&
  McCullough}]{MarchRussell:2008dy}
March-Russell, J., McCabe, C., \& McCullough, M. 2009, JHEP, 05, 071

\bibitem[{Millar {et~al.}(2017)Millar, Redondo, \& Steffen}]{Millar:2017eoc}
Millar, A.~J., Redondo, J., \& Steffen, F.~D. 2017, JCAP, 1710, 006

\bibitem[{{Myeong} {et~al.}(2018){Myeong}, {Evans}, {Belokurov}, {Sanders}, \&
  {Koposov}}]{2018ApJ...856L..26M}
{Myeong}, G.~C., {Evans}, N.~W., {Belokurov}, V., {Sanders}, J.~L., \&
  {Koposov}, S.~E. 2018, \apjl, 856, L26

\bibitem[{Myeong {et~al.}(2018)Myeong, Evans, Belokurov, Sanders, \&
  Koposov}]{Myeong:2018kfh}
Myeong, G.~C., Evans, N.~W., Belokurov, V., Sanders, J.~L., \& Koposov, S.~E.
  2018, arXiv:1805.00453

\bibitem[{Necib {et~al.}(2018)Necib, Lisanti, Garrison-Kimmel, Wetzel,
  Sanderson, Hopkins, Faucher-Gigu{\`e}re, \& Kere{\v s}}]{Necib:2018}
Necib, L., Lisanti, M., Garrison-Kimmel, S., {et~al.} 2018, arXiv:1810.12301

\bibitem[{{Oort}(1932)}]{1932BAN.....6..249O}
{Oort}, J.~H. 1932, \bain, 6, 249

\bibitem[{{Ostriker} {et~al.}(1974){Ostriker}, {Peebles}, \&
  {Yahil}}]{1974ApJ...193L...1O}
{Ostriker}, J.~P., {Peebles}, P.~J.~E., \& {Yahil}, A. 1974, \apjl, 193, L1

\bibitem[{P\'erez \& Granger(2007)}]{PER-GRA:2007}
P\'erez, F., \& Granger, B.~E. 2007, Computing in Science and Engineering, 9,
  21

\bibitem[{{Pillepich} {et~al.}(2014){Pillepich}, {Kuhlen}, {Guedes}, \&
  {Madau}}]{Pillepich:2014784}
{Pillepich}, A., {Kuhlen}, M., {Guedes}, J., \& {Madau}, P. 2014, \apj, 784,
  161

\bibitem[{{Pont} {et~al.}(1994){Pont}, {Mayor}, \&
  {Burki}}]{1994A&A...285..415P}
{Pont}, F., {Mayor}, M., \& {Burki}, G. 1994, \aap, 285, 415

\bibitem[{Price-Whelan {et~al.}(2017)Price-Whelan, Sipocz, Major, \&
  Oh}]{PriceWhelan:2017}
Price-Whelan, A., Sipocz, B., Major, S., \& Oh, S. 2017, {adrn/gala}, v.0.2.1,
  Zenodo, doi:10.5281/zenodo.833339

\bibitem[{Purcell {et~al.}(2011)Purcell, Bullock, Tollerud, Rocha, \&
  Chakrabarti}]{Purcell:2011nf}
Purcell, C.~W., Bullock, J.~S., Tollerud, E., Rocha, M., \& Chakrabarti, S.
  2011, Nature, 477, 301

\bibitem[{Read(2014)}]{Read:2014qva}
Read, J.~I. 2014, J. Phys., G41, 063101

\bibitem[{Schlegel {et~al.}(1998)Schlegel, Finkbeiner, \&
  Davis}]{Schlegel:1997yv}
Schlegel, D.~J., Finkbeiner, D.~P., \& Davis, M. 1998, Astrophys. J., 500, 525

\bibitem[{Sch{\"o}nrich(2012)}]{doi:10.1111/j.1365-2966.2012.21631.x}
Sch{\"o}nrich, R. 2012, Monthly Notices of the Royal Astronomical Society, 427,
  274

\bibitem[{{Sch{\"o}nrich} \& {Binney}(2012)}]{2012MNRAS.419.1546S}
{Sch{\"o}nrich}, R., \& {Binney}, J. 2012, \mnras, 419, 1546

\bibitem[{Sloan {et~al.}(2016)}]{Sloan:2016aub}
Sloan, J.~V., {et~al.} 2016, Phys. Dark Univ., 14, 95

\bibitem[{Sloane {et~al.}(2016)Sloane, Buckley, Brooks, \&
  Governato}]{Sloane:2016kyi}
Sloane, J.~D., Buckley, M.~R., Brooks, A.~M., \& Governato, F. 2016,
  arXiv:1601.05402

\bibitem[{Smith {et~al.}(2009)Smith, Evans, Belokurov, Hewett, Bramich,
  Gilmore, Irwin, Vidrih, \& Zucker}]{Smith:2009kr}
Smith, M.~C., Evans, N.~W., Belokurov, V., {et~al.} 2009, Mon. Not. Roy.
  Astron. Soc., 399, 1223

\bibitem[{Venn {et~al.}(2004)Venn, Irwin, Shetrone, Tout, Hill, \&
  Tolstoy}]{2004AJ....128.1177V}
Venn, K.~A., Irwin, M., Shetrone, M.~D., {et~al.} 2004, \aj, 128, 1177

\bibitem[{{Vergados} \& {Semertzidis}(2017)}]{2017NuPhB.915...10V}
{Vergados}, J.~D., \& {Semertzidis}, Y.~K. 2017, Nuclear Physics B, 915, 10

\bibitem[{{Vogelsberger} \& {White}(2011)}]{2011MNRAS.413.1419V}
{Vogelsberger}, M., \& {White}, S.~D.~M. 2011, \mnras, 413, 1419

\bibitem[{Vogelsberger {et~al.}(2009)Vogelsberger, Helmi, Springel, White,
  Wang, Frenk, Jenkins, Ludlow, \& Navarro}]{Vogelsberger:2008qb}
Vogelsberger, M., Helmi, A., Springel, V., {et~al.} 2009, Mon. Not. Roy.
  Astron. Soc., 395, 797

\bibitem[{{Wang} {et~al.}(2011){Wang}, {Navarro}, {Frenk}, {White}, {Springel},
  {Jenkins}, {Helmi}, {Ludlow}, \& {Vogelsberger}}]{2011MNRAS.413.1373W}
{Wang}, J., {Navarro}, J.~F., {Frenk}, C.~S., {et~al.} 2011, \mnras, 413, 1373

\bibitem[{White \& Rees(1978)}]{White:1977jf}
White, S. D.~M., \& Rees, M.~J. 1978, Mon. Not. Roy. Astron. Soc., 183, 341

\bibitem[{Yanny {et~al.}(2009)}]{Yanny:2009kg}
Yanny, B., {et~al.} 2009, Astron. J., 137, 4377

\bibitem[{Zemp {et~al.}(2009)Zemp, Diemand, Kuhlen, Madau, Moore,
  {et~al.}}]{Zemp:2008gw}
Zemp, M., Diemand, J., Kuhlen, M., {et~al.} 2009, Mon. Not. Roy. Astron. Soc.,
  394, 641

\end{thebibliography}

\newpage

\onecolumngrid

\newpage
\appendix

\setcounter{equation}{0}
\setcounter{figure}{0}
\setcounter{table}{0}
\setcounter{section}{0}
\makeatletter
\renewcommand{\theequation}{S\arabic{equation}}
\renewcommand{\thefigure}{S\arabic{figure}}
\renewcommand{\thetable}{S\arabic{table}}

In this Appendix, we include some additional figures to supplement the discussion in the main text.  We provide residual maps that demonstrate the quality of the model fit, show how the results in the main text vary in five radial bins from $r=7.5$ to 10~kpc, and provide corner plots for the halo, disk, and substructure parameters. \vspace{0.6in}

\begin{figure*}[h]
\centering
\includegraphics[width=0.75\textwidth]{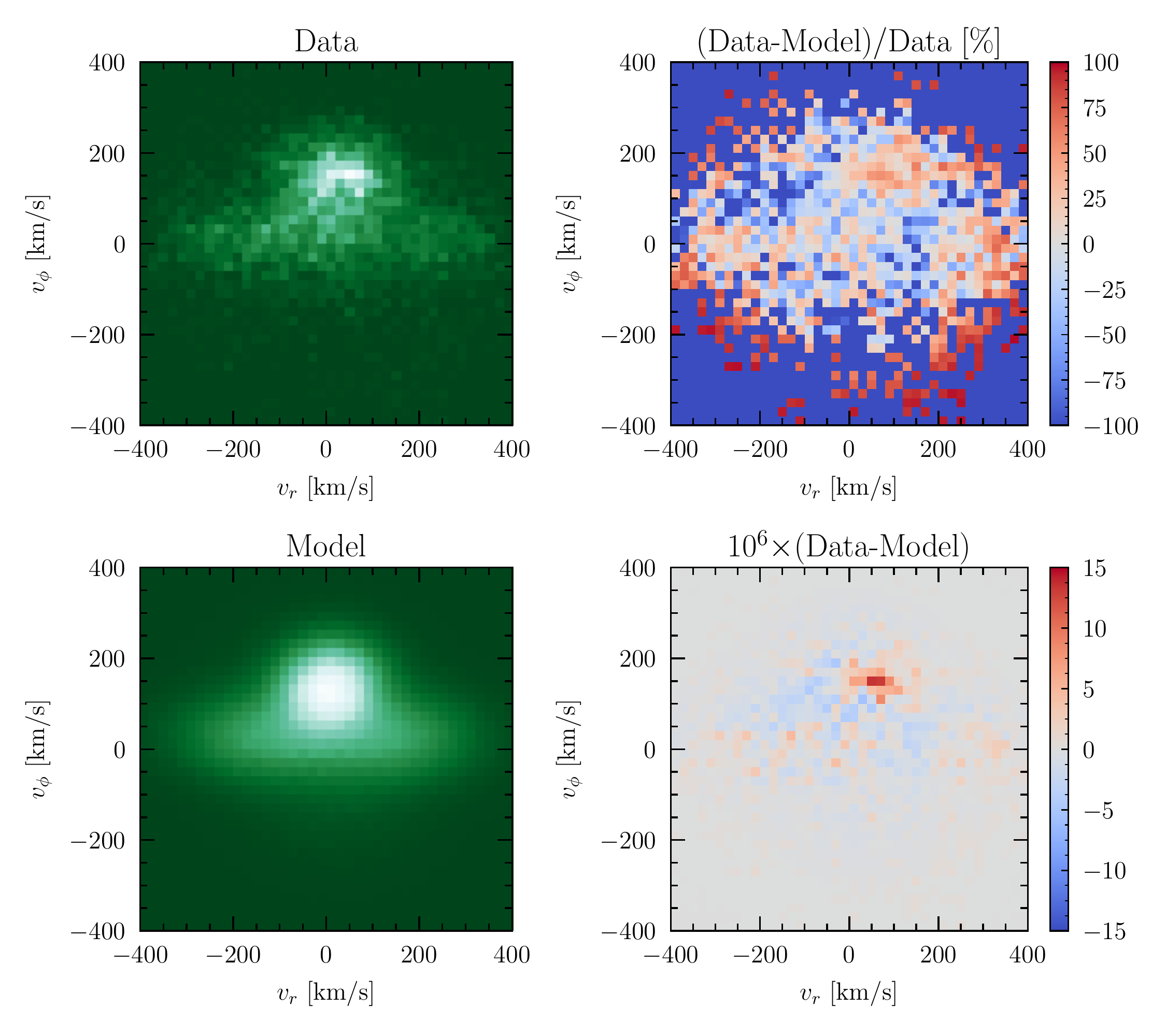}
\caption{A comparison of the data and best-fit model distributions in the $v_r-v_\phi$ plane for the SDSS-\emph{Gaia}~DR2 data in the region $r\in[7.5, 8.5]$~kpc and $|z| > 2.5$~kpc.  The left column shows a count map of the data (top) and model (bottom).  The right column compares the two explicitly, showing the fractional (top) and total (bottom) residuals.  The bottom right plot is the difference in the value of the binned histograms, where both the data and model are normalized to unity beforehand.}
\label{fig:residual}
\end{figure*}

\begin{figure*}[t] 
\centering
\includegraphics[width=0.75\textwidth]{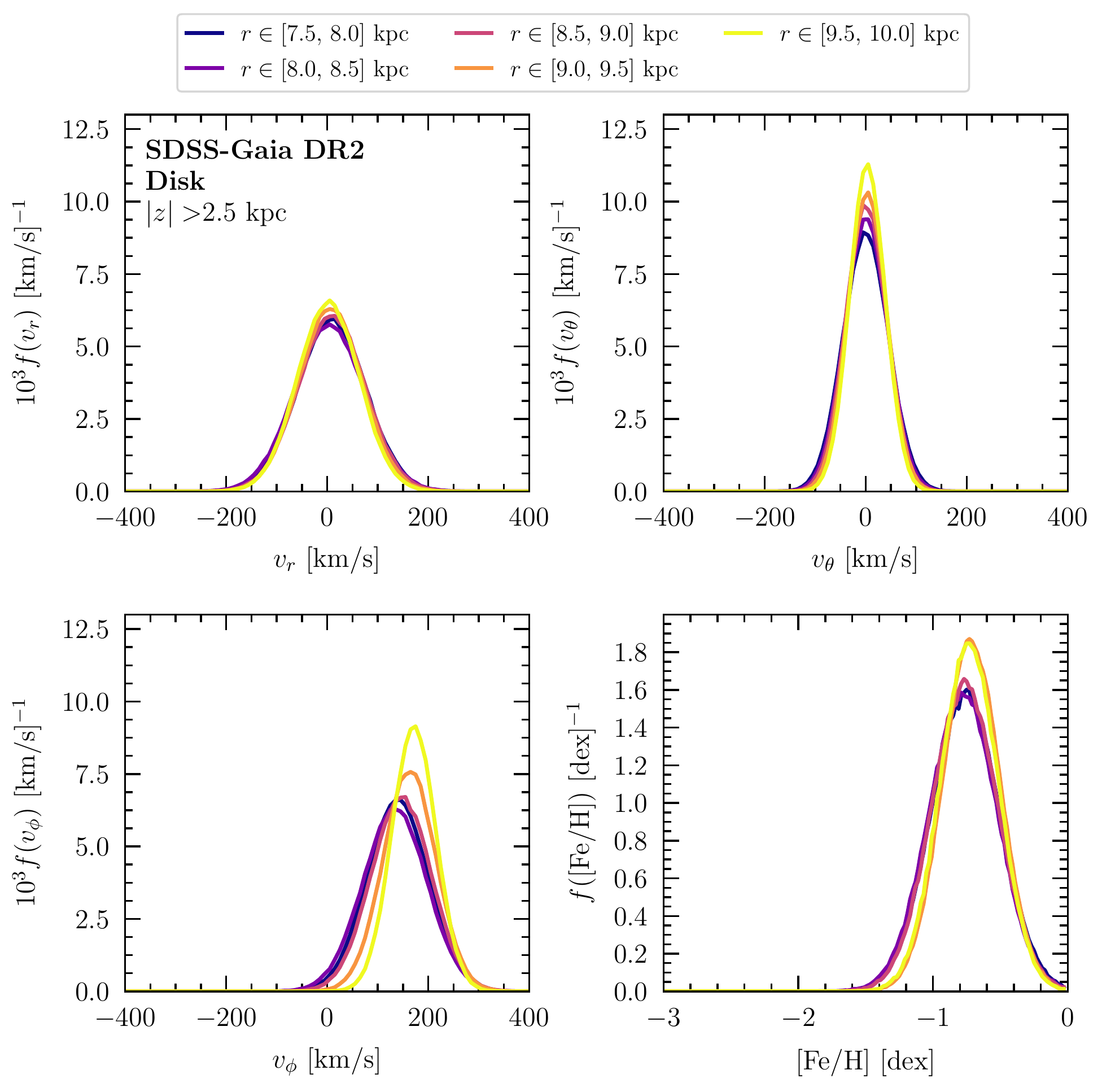}
\caption{The same as Fig.~\ref{fig:zcutDisk}, except for different radial cuts and fixing $|z|>2.5$~kpc.}
\label{fig:Subs_rgc3kpc}
\end{figure*}

\begin{figure*}[t] 
\centering
\includegraphics[width=0.75\textwidth]{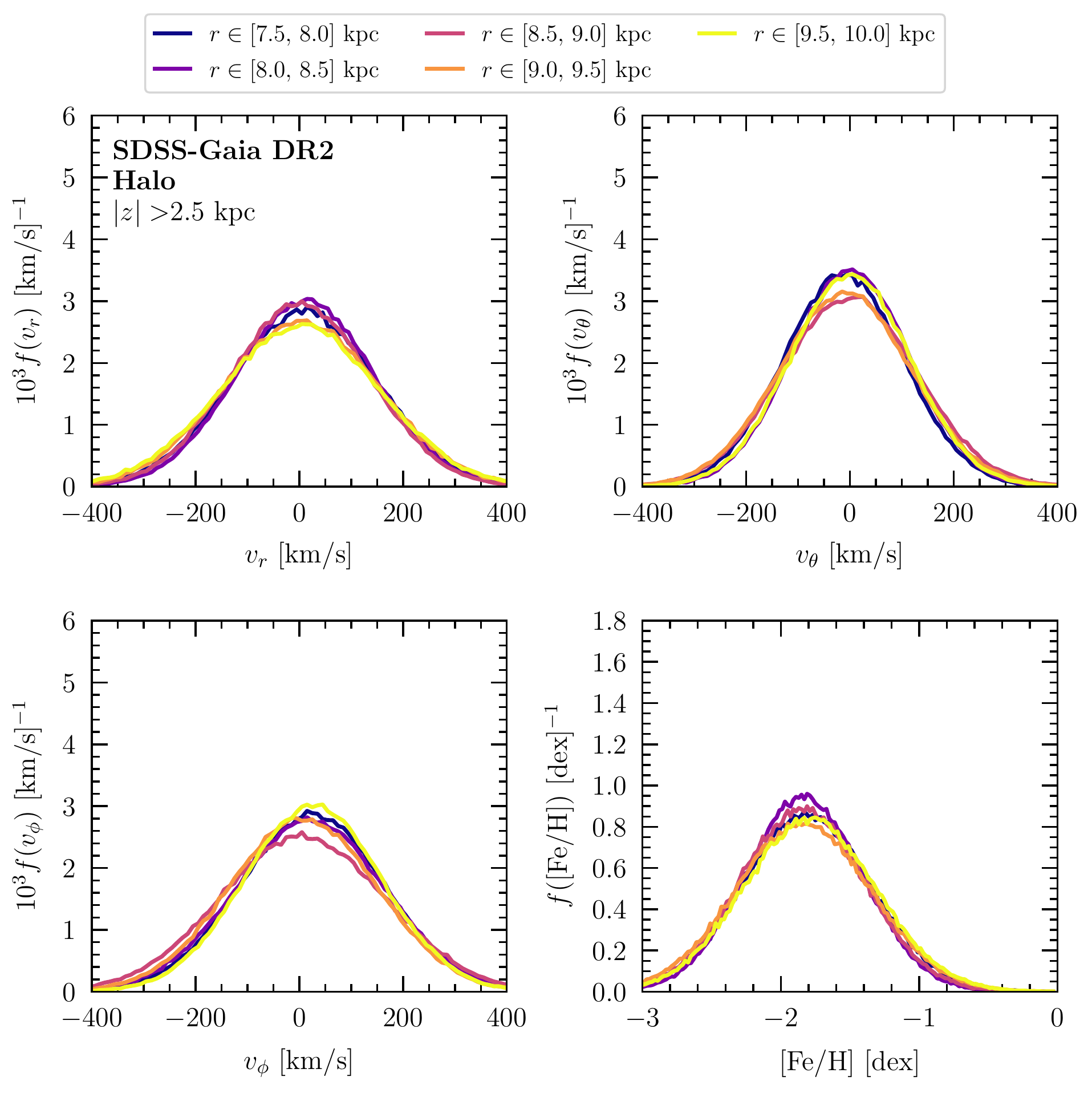}
\caption{The same as Fig.~\ref{fig:zcutHalo}, except for different radial cuts.}
\label{fig:Subs_rgc3kpc}
\end{figure*}

\begin{figure*}[t] 
\centering
\includegraphics[width=0.75\textwidth]{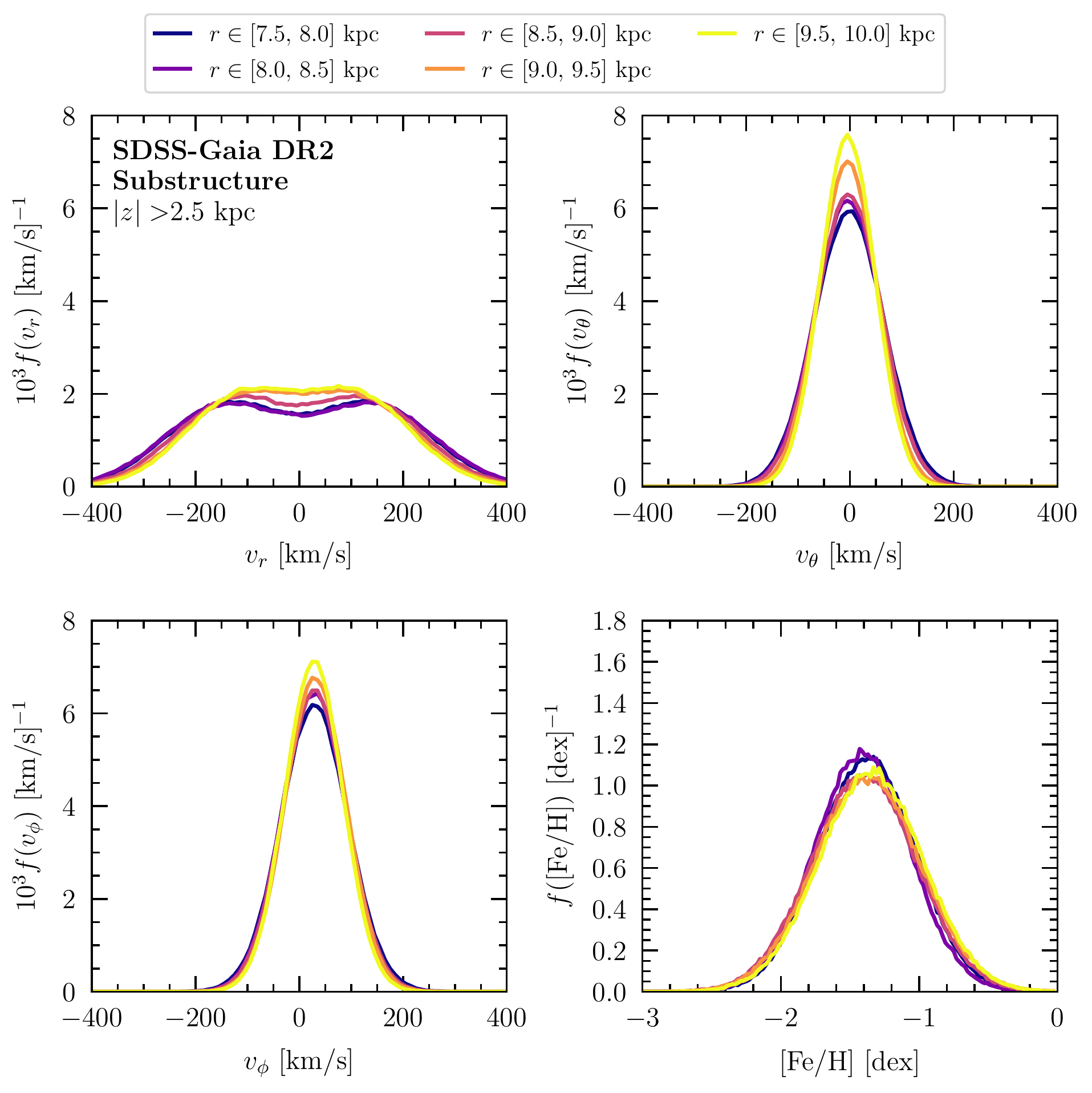}
\caption{The same as Fig.~\ref{fig:zcutSubs}, except for different radial cuts.}
\label{fig:Subs_rgc3kpc}
\end{figure*}

\begin{figure*}[t] 
\centering
\includegraphics[width=0.75\textwidth]{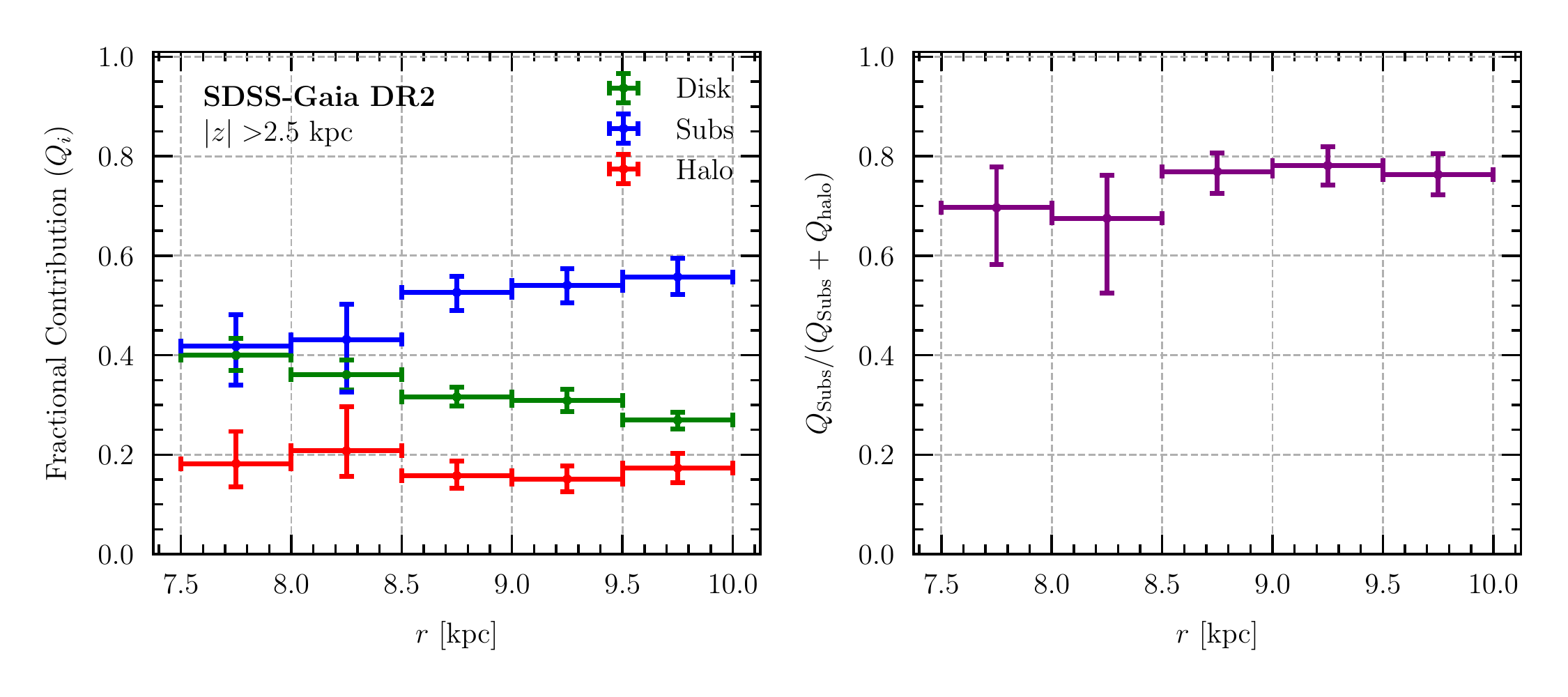}
\caption{The same as Fig.~\ref{fig:fractions}, except for different radial cuts.}
\label{fig:fractions_rgc}
\end{figure*}


\begin{figure*}[t] 
\centering
\includegraphics[width=0.95\textwidth]{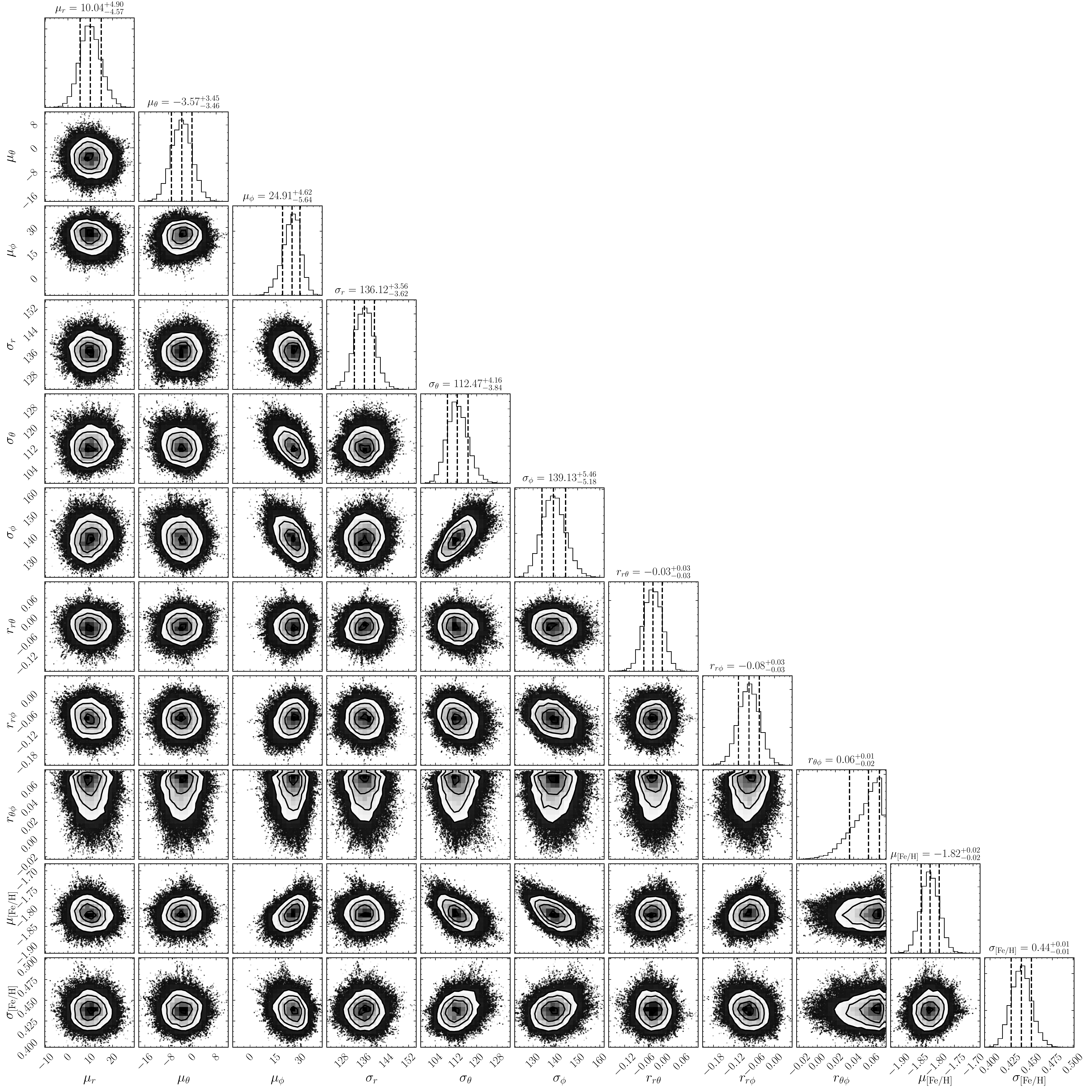}
\caption{Corner plot for the halo model parameters in the region $r\in[7.5, 8.5]$ kpc, and $|z|>2.5$ kpc.}
\label{fig:corner_halo}
\end{figure*}

\begin{figure*}[t] 
\centering
\includegraphics[width=0.95\textwidth]{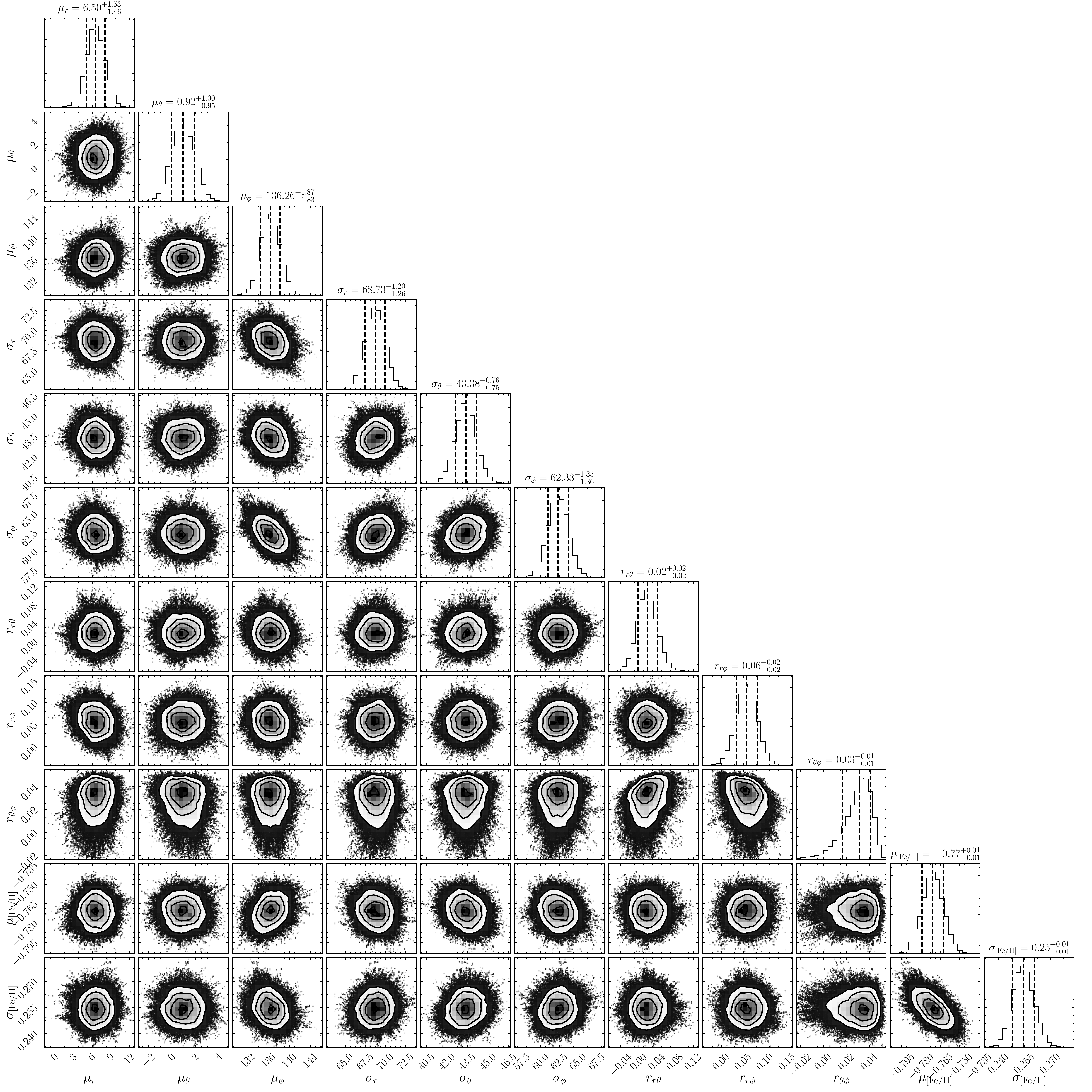}
\caption{Corner plot for the disk model parameters in the region $r\in[7.5, 8.5]$ kpc, and $|z|>2.5$ kpc.}
\label{fig:corner_disk}
\end{figure*}

\begin{figure*}[t] 
\centering
\includegraphics[width=0.95\textwidth]{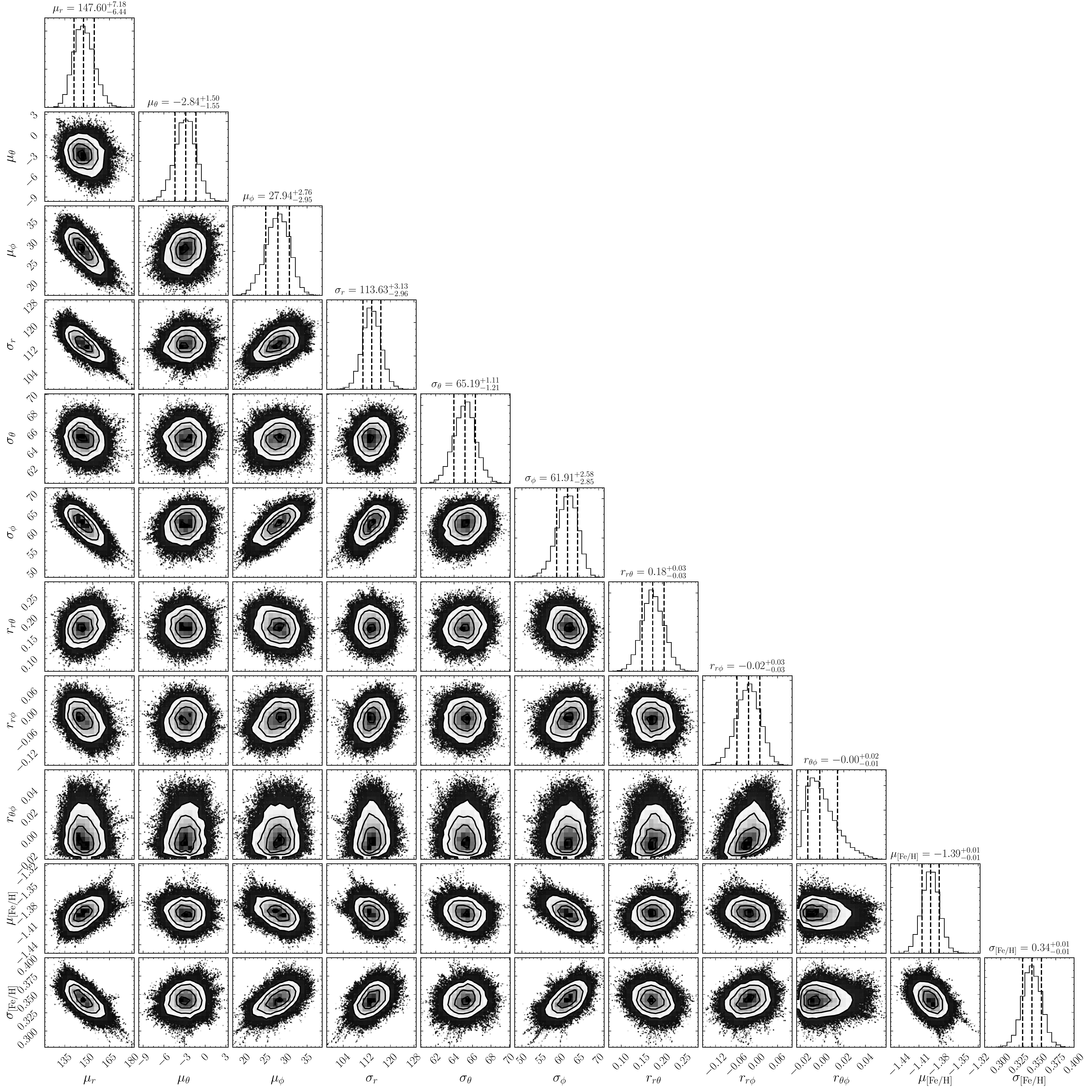}
\caption{Corner plot for the substructure model parameters in the region $r\in[7.5, 8.5]$ kpc, and $|z|>2.5$ kpc.}
\label{fig:corner_lobes}
\end{figure*}

\end{document}